\documentclass[aps,eqsecnum,prb,twocolumn]{revtex4}
\usepackage{graphicx}
\usepackage{dcolumn}
\usepackage{bm}

\begin{document}
\title{
  Troubleshooting Time-Dependent Density-Functional 
  Theory for Photochemical Applications: Oxirane
}

\author{Felipe Cordova, L. \surname{Joubert Doriol}, Andrei Ipatov, and Mark E. Casida}
\email[]{Mark.Casida@UJF-Grenoble.Fr}
\affiliation{
        Laboratoire de Chimie Th\'eorique,
        D\'epartement de Chimie Mol\'ecularie (DCM, UMR CNRS/UJF 5250),
        Institut de Chimie Mol\'eculaire de Grenoble (ICMG, FR2607),
        Universit\'e Joseph Fourier (Grenoble I),
        301 rue de la Chimie, BP 53,
        F-38041 Grenoble Cedex 9, France}

\author{Claudia Filippi}
\email[]{filippi@lorentz.leidenuniv.nl}
\affiliation{
       Instituut-Lorentz for Theoretical Physics,
       Universiteit Leiden, Niels Bohrweg 2, Leiden, NL-2333 CA, The
       Netherlands}

\author{Alberto Vela}
\affiliation{
        Departamento de Qu\'{\i}mica, Cinvestav,
        Avenida Instituto Polit\'ecnico Nacional 2508, 
        A.P. 14-740 Mexico D.F. 07000,
        Mexico}


  
\begin{abstract}

The development of analytic-gradient methodology for excited states within
conventional time-dependent density-functional theory (TDDFT) would seem to
offer a relatively inexpensive alternative to better established quantum-chemical 
approaches for the modeling of photochemical reactions. However, even 
though TDDFT is formally exact, practical calculations involve the use of approximate 
functionals, in particular the TDDFT adiabatic approximation, whose use in 
photochemical applications must be further validated.  Here, we investigate the 
prototypical case of the symmetric CC ring opening of oxirane.  We demonstrate
by direct comparison with the results of high-quality quantum Monte Carlo calculations
that, far from being an approximation on TDDFT,  the Tamm-Dancoff approximation 
(TDA) is a practical necessity for avoiding triplet instabilities and singlet near 
instabilities, thus helping maintain energetically reasonable excited-state 
potential energy surfaces during bond breaking.  Other difficulties one would 
encounter in modeling oxirane photodynamics are pointed out but none of these 
is likely to prevent a qualitatively correct TDDFT/TDA description of photochemistry 
in this prototypical molecule.

\end{abstract}

\maketitle

\section{Introduction}
\label{sec:intro}

A complete understanding of photochemistry often requires photodynamics calculations
since photochemical reactions may have excess energy and do not follow the 
lowest energy pathway, and since dynamic, as much as energetic, considerations 
often govern how a reaction jumps from one electronic potential energy surface 
(PES) to another.  Unfortunately, full photodynamics calculations are prohibitively 
expensive for all but the smallest molecules unless simplifying approximations 
are made.  One such approximation is to adopt a Tully-type mixed quantum/classical 
surface-hopping (SH) trajectory approach \cite{TP71,T90}
where the electrons are described quantum mechanically and the nuclei classically. 
Then, since the quantum part of the calculation remains the computational bottleneck, 
further simplifications are required for an efficient electronic structure
computation of the photochemical dynamics.

Time-dependent density-functional theory (TDDFT) \cite{MUN+06} represents a promising
and relatively inexpensive alternative \cite{DM02} to accurate but more costly quantum 
chemical approaches for on-the-fly calculations of excitation energies. 
TDDFT methods for excited-state dynamics are not yet a standard part of the 
computational chemistry repertory but are coming on line. In particular, the basic 
methodology for the computation of analytic gradients has now been implemented
in a number of codes \cite{VA99,VA00,FA02,RF05,H03,DK05} and important 
practical progress is being made on the problem of calculating SH integrals within
TDDFT \cite{CM00,B02,CDP05,M06,HCP06,TTR07}.   Indeed first applications of
mixed TDDFT/classical SH dynamics are being reported \cite{CDP05,HCP06,TTR07}.
Problems in this new technology can be expected and
modifications in the basic TDDFT methods will be necessary 
to enlarge the class of its potential applications. 
In particular, practical TDDFT calculations employ approximate functionals whose performance
in describing photochemical problems requires careful validation.

Few applications of TDDFT to problems relevant to photochemistry can be found
in the literature~\cite{DKZ01a,DKZ01b,DKSZ02,SDK+02,OGB02,CSM+06} and the few 
attempts made to assess the value of TDDFT for photochemical applications
report either encouraging \cite{CCS98,FMO04,BGDF06} or discouraging 
results \cite{BM00,CTR00,RCT04,WGB+04,LQM06}.
The fundamental reasons for these descrepencies stem from the fact that 
conventional TDDFT has several problems which may or may not be fatal for 
modeling a given photochemical reaction.  The main difficulties encountered
in conventional TDDFT include the underestimation of the ionization threshold
\cite{CCS98}, the underestimation of charge transfer excitations 
\cite{TAH+99,CGG+00,DWH03}, and the lack of explicit two- and higher-electron
excitations \cite{C95,MZCB04,CZMB04,C05}.  Still other difficulties are discussed in
pertinant reviews \cite{C01,MBA+01,ORR02,D03,MWB03,SM05,MUN+06}.

To help TDDFT become a reliable tool for modeling photochemistry, we must 
first obtain a clear idea of what are its most severe problems. It is the 
objective of this article to identify the most critical points where improvement 
needs to be made, at least for one prototypical molecule and one type of reaction path.
The reaction chosen for our study is the photochemical ring opening of oxirane.  
The photochemistry of oxirane and of oxirane derivatives has been much studied
both theoretically and experimentally  (Appendix~\ref{sec:photochem}).
It is a text-book molecule used to discuss orbital control of stereochemistry
in the context of electrocyclic ring opening.  Textbook discussions usually
focus on CC ring opening (see e.g.\ Ref.~\cite{MB90} pp.\ 258-260) but a discussion
of CO ring opening may also be found (Ref.~\cite{MB90} pp.\ 408-411).
At first glance, this would seem to be a particularly good test case for TDDFT.
That is, oxirane is a simple molecule with a relatively
simple photochemistry where TDDFT should work and, if not, where its failures will 
be particularly easy to analyze.  
In this paper, we will focus on the symmetric ring opening of this molecule even
though the photochemical ring opening does not follow a symmetric pathway.
The reason for our choice is the usual reason \cite{symmetry}, namely that the use of
symmetry greatly facilitates analysis and hence the construction of and comparison
with highly accurate quantum Monte Carlo results.  A mixed quantum/classical
SH trajectory study of asymmetric ring opening will be reported elsewhere \cite{T07}.

%

In the next section, we give a brief review of DFT and of TDDFT.  In sec.~\ref{sec:QMC},
we review the formalism behind the more exact theory against which we will be 
comparing our TDDFT results.  Computational details are given in Sec.~\ref{sec:details}.
Section~\ref{sec:results} reports our results and discussion and Sec.~\ref{sec:conclude}
summarizes.
  
\section{(Time-Dependent) Density-Functional Theory}
\label{sec:TDDFT}

The (TD)DFT PES for the $I$th excited state
is calculated at each nuclear configuration by
adding the $I$th TDDFT excitation energy to the 
DFT ground state energy, so
$E_I  = E_0 + \omega_I$.  [Hartree atomic units
($\hbar=m=e=1$) are used throughout the
paper.]  Even though the use of parentheses in the
expression (TD)DFT better emphasizes the hybrid 
DFT + TDDFT nature of the calculation, we will usually follow
the common practice of simply refering 
to TDDFT PESs.   The purpose of this section is to
review those aspects of TDDFT most necessary for
understanding the rest of the paper.

The hybrid DFT + TDDFT nature of (TD)DFT calculations 
indicates that  problems with the ground state PES
can easily become problems for the excited-state PESs.
Thus it is important to begin with a few words about
the ground-state problem.  The simplest methods for treating the
ground state are the Hartree-Fock (HF) method and the Kohn-Sham 
formulation of DFT.  
In recent years, the use of hybrid functionals has permitted the two
energy expressions to be written in the same well-known form,
\begin{eqnarray}
E^{\text{hybrid}}_{xc} = E^{\text{GGA}}_{xc} + c_x ( E^{\text{HF}}_x - E^{\text{GGA}}_x )
\end{eqnarray}
where $E^{\text{HF}}_x$ is the HF exchange energy and $E^{\text{GGA}}_x$ and
$E^{\text{GGA}}_{xc}$ are generalized gradient approximation (GGA) exchange (x) and
exchange-correlation (xc) energies.
The coefficient $c_x$ controls the amount of Hartree-Fock exchange, being unity for
Hartree-Fock, zero for pure DFT, and fractional (typically around 0.25 \cite{PEB96})
for hybrid functionals.  For more information about DFT, we refer the reader 
to Refs.~\cite{PY89,DG90,KH00}.

Time-dependent density-functional theory offers a rigorous approach to
calculating excitation energies.  
Unlike time-dependent Hartree-Fock (TDHF), TDDFT is formally exact.  
Consequently, although approximate exchange-correlation functionals must
be used in practice, we can hope that good approximations will lead to better results than
those obtained from TDHF which completely lacks correlation effects.
In practice, the use of hybrid functionals means that TDDFT contains TDHF as a 
particular choice of functional (i.e., the exchange-only hybrid functional with $c_x=1$).

The most common implementation of TDDFT is via a Kohn-Sham formalism using the
so-called adiabatic approximation which assumes that the self-consistent field
responds instantaneously and without memory to any temporal change in the charge
density.  This allows the time-dependent exchange-correlation action quantity in
TDDFT to be replaced with the more familiar exchange-correlation energy from 
conventional TDDFT.  Excitations may be obtained from the linear response formulation
of TDDFT (LR-TDDFT).  A key quantity in LR-TDDFT is the exchange-correlation kernel,
\begin{equation}
  f_{xc}^{\sigma,\tau}({\bf r},{\bf r}')
  = \frac{
          \delta^2 E_{xc}[\rho_\uparrow,\rho_\downarrow]
    }
    {
          \delta \rho_\sigma({\bf r}) \delta \rho_\tau({\bf r}')
    } \, ,
   \label{eq:TDDFT.8}
\end{equation}
which, along with the Hartree kernel, $ f_{H}^{\sigma,\tau}({\bf r},{\bf r}') 
= 1/ \left| {\bf r} - {\bf r}' \right| $,
and the Hartree-Fock exchange kernel (whose integrals can be written in terms of the 
Hartree kernel) determines the linear response of the Kohn-Sham self-consistent field 
in the adiabatic approximation.  In Casida's formulation \cite{C95}, the excitation energies
are obtained by solving a random phase approximation (RPA)-like pseudo-eigenvalue equation,
\begin{equation}
   \left[ \begin{array}{cc} {\bf A} & {\bf B} \\
         {\bf B} & {\bf A} \end{array} \right] 
   \left( \begin{array}{c} \vec{X}_I \\ \vec{Y}_I \end{array} \right) 
   = \omega_I
   \left[ \begin{array}{cc} +{\bf 1} & {\bf 0} \\
         {\bf 0} & -{\bf 1} \end{array} \right] 
   \left( \begin{array}{c} \vec{X}_I \\ \vec{Y}_I \end{array} \right) \, , 
   \label{eq:TDDFT.10}
\end{equation}
where the matrices ${\bf A}$ and ${\bf B}$ are defined by,
\begin{eqnarray}
  A_{ia\sigma,jb\tau} & = & \delta_{i,j} \delta_{a,b} \delta_{\sigma,\tau} 
  \left( \epsilon_{a\sigma} - \epsilon_{i\sigma} \right) 
  + \left( ia \vert f_H \vert bj \right) +  
  \left( ia  \vert f_{xc}^{\sigma,\tau} \vert bj  \right) \nonumber \\
  & - & c_x  \delta_{\sigma,\tau} \left[ 
   \left( ij \vert f_H \vert ba \right) +
  \left( ia  \vert f_{x}^{\sigma,\sigma} \vert bj  \right) \right] \nonumber \\
  B_{ia\sigma,bj\tau} & = &  \left( ia  \vert f_H \vert jb  \right)
   +  \left( ia  \vert f_{xc}^{\sigma,\tau} \vert jb  \right)    
  \nonumber \\
  & - & c_x \delta_{\sigma,\tau} \left[ 
  \left( ib \vert f_H \vert ja\right) +
  \left( ia  \vert f_{x}^{\sigma,\sigma} \vert jb  \right)
   \right] \, , 
  \label{eq:TDDFT.11}
\end{eqnarray} 
and the integrals are in Mulliken charge cloud notation,
\begin{equation}
  \left( pq \vert f \vert rs \right) = \int \int \psi_p^* ({\bf r}) \psi_q({\bf r}) 
  f({\bf r}, {\bf r}') \psi_r^* ({\bf r}') \psi_s({\bf r}') \, d{\bf r} d{\bf r}' \, .
  \label{eq:TDDFT.12}
\end{equation}

It is to be emphasized that the TDDFT adiabatic approximation includes only dressed
one-electron excitations \cite{C95,C05}.  This is particularly easy to see in the
context of the Tamm-Dancoff approximation (TDA) to Eq.~(\ref{eq:TDDFT.10})
whose TDDFT variant is usually attributed to Hirata and Head-Gordon \cite{HH99}.
The TDA simply consists of neglecting the ${\bf B}$ matrices to obtain,
$ {\bf A} \vec{X}_I = \omega_I \vec{X}_I $.
The number of possible solutions to this equation is the dimensionality of ${\bf A}$
which is just the number of single excitations.  In fact, the LR-TDHF TDA (exchange-only
hybrid functional with $c_x=1$) 
is simply the well-known configuration interaction singles (CIS) method. 

An issue of great importance in the context of the present work is triplet instabilities.
These have been first analyzed by Bauernschmitt and Ahlrichs \cite{BA96} in the context of
DFT, but the explicit association with LR-TDDFT excitation energies was made later 
\cite{CGG+00,C01}.
Following Ref.~\cite{C01}, we suppose that the ground-state DFT calculation has been
performed using a  same-orbitals-for-different-spin (SODS) ansatz and we now wish 
to test to see if releasing the SODS restriction to give a 
different-orbitals-for-different-spin (DODS) solution will lower the energy.
To do so we consider an arbitary unitary tranformation of the orbitals,
\begin{equation}
  \psi_{r\sigma}^\lambda ({\bf r}) = \exp\left[i\lambda \left(\hat{R}+i\hat{I}\right)\right]
  \psi_{r\sigma}({\bf r})\,.
\end{equation}
where $\hat{R}$ and $\hat{I}$ are real operators.  The corresponding energy
expression is,
\begin{equation}
  E_\lambda = E_0 + \lambda^2 \left[ \vec{R}^\dagger \left( {\bf A} - {\bf B} \right)
  \vec{R} + \vec{I}^\dagger \left(  {\bf A} + {\bf B} \right) \vec{I} \right]
  + {\cal O}(\lambda^3) \, ,
  \label{eq:TDDFT.15}
\end{equation}
where matrix elements of the $\hat{R}$ and $\hat{I}$ operators have been arranged in
column vectors and the ${\cal O}(\lambda)$ term disappears because the energy has already
been minimized before considering symmetry-breaking.  The presence of the terms 
$({\bf A} \pm {\bf B})$ shows the connection with the pseudoeigenvalue 
problem~(\ref{eq:TDDFT.10}) which can be rewritten as the eigenvalue problem,
\begin{equation}
 \left( {\bf A} + {\bf B} \right) \left( {\bf A} - {\bf B} \right) \vec{Z}_I
  = \omega_I^2 \vec{Z}_I\,,
 \label{omega2_eq}
\end{equation}
where $\vec{Z}_I=\vec{X}_I-\vec{Y}_I$.
It is an easy consequence of Eq.~(\ref{eq:TDDFT.11}) that the matrix 
$\left( {\bf A} - {\bf B} \right)$ is always positive definite for pure functionals ($c_x=0$),
as long as the {\em aufbau} principle is obeyed.  However $\left( {\bf A} + {\bf B} \right)$ 
may have negative eigenvalues.  In that case, the energy $E_\lambda$ will fall below
$E_0$ for some value of $\vec{I}$, indicating a lower symmetry-broken solution.  
At the same time, this will correspond to a negative value of $\omega_I^2$ 
(i.e., an imaginary value of $\omega_I$.) A further analysis shows the 
corresponding eigenvector corresponds to a triplet
excitation, hence the term ``triplet instability'' for describing this phenomenon.
In principle, when hybrid functionals are used, singlet instabilities are 
also possible.

Thus imaginary excitation energies in LR-TDDFT are an indication that
something is wrong in the description of the ground state whose time-dependent response
is being used to obtain those excitation energies.  As pointed out in Refs.~\cite{CGG+00}
and~\cite{CIC06}, the TDA actually acts to {\em decouple} the excited-state 
problem from the ground-state problem so that TDA excitations may actually
be {\em better} than full LR-TDDFT ones.  Typically, full and TDA
LR-TDDFT excitation energies are close near a molecule's equilibrium geometry, where
a single-determinantal wave function is a reasonable first approximation and begin to
differ as the single-determinantal description breaks down.

Interestingly, there is a very simple argument against symmetry breaking in exact 
DFT for molecules with a nondegenerate ground state. Since
the exact ground-state wave function of these molecules is a singlet belonging 
to the totally symmetric representation and so with the same symmetry as the molecule,
the spin-up and spin-down charge 
densities will be equal and also have the same symmetry as the molecule.
It follows that the same must hold for the spin-up and spin-down
components of the exact exchange-correlation potential.
Since the potentials are the same, then the molecular orbitals must also be the same for 
different spins.  Thus, no symmetry breaking is expected in exact DFT for molecules
with a nondegenerate ground state.  
Note however that the functional is still dependent
on spin so that the spin kernels $ f_{xc}^{\uparrow,\uparrow}({\bf r},{\bf r}')$ and 
$f_{xc}^{\uparrow,\downarrow}({\bf r},{\bf r}')$ are different.

The interested reader will find additional information about TDDFT in the recent book of Ref.~\cite{MUN+06}.
  
\section{Quantum Monte Carlo}
\label{sec:QMC}

The quality of our (TD)DFT calculations is judged against highly accurate
quantum Monte Carlo (QMC) results. For each state of interest, the QMC results
are obtained in a three step procedure.  First, a conventional complete active space 
(CAS) self-consistent field (SCF) calculation is performed. The resultant CASSCF 
wave function is then reoptimized in the presence of a Jastrow 
factor to include dynamical correlation and used in a variational Monte Carlo 
(VMC) calculation.  Finally, the VMC result is further improved via diffusion 
Monte Carlo (DMC).

To properly describe a reaction involving bond breaking, it is necessary
to adopt a multi-determinantal description of the wave function. In the CASSCF method, 
a set of active orbitals is selected, whose occupancy is allowed to vary, while all
other orbitals are fixed as either doubly occupied or unoccupied.  In a CASSCF($n$,$m$) 
calculation, $n$ electrons are distributed among an active space of $m$ orbitals and 
all possible resulting space- and spin-symmetry-adapted configuration state functions (CSFs)
are constructed.  The final CASSCF($n$,$m$) wave function consists of a linear 
combination of these CSFs, like in a full configuration interaction (CI) calculation for 
$n$ electrons in $m$ orbitals, except that also the orbitals are now
optimized to minimize the total energy.  
In the present article, we consider 4 electrons 
in an active space of 6 orbitals which represents the minimal active space for a proper
description of all 9 states of interest.  

When several states of the same symmetry are requested, there is a danger 
in optimizing the higher states that 
their energy is lowered enough
to approach and mix with lower states, thus giving an unbalanced
description of excitation energies.  A well-established solution to this 
problem is the use of a state averaged (SA) CASSCF approach where the weighted 
average of the energies of the states under consideration is optimized~\cite{WM81,HJO00}. 
The wave functions of the different states depend on their individual sets of 
CI coefficients using a common set of orbitals. Orthogonality is ensured via 
the CI coefficients and a generalized variational theorem applies. Obviously, 
the SA-CASSCF energy of the lowest state will be higher than the CASSCF energy 
obtained without SA. In the present work, we need to apply the SA procedure 
only for the $2^1A_1$ state in the $C_{2v}$ ring opening.  Although the CASSCF 
and SA-CASSCF $1^1A_1$ energies are found to be really very close, we calculate 
the $2^1A_1$ energy as
$ E_{\text{CAS}}(1^1A_1)
   + \left[ E_{\text{CAS}}^{\text{SA}}(2^1A_1)- E_{\text{CAS}}^{\text{SA}}(1^1A_1) \right]$.
A similar procedure is used in the case of the corresponding QMC calculations.

QMC methods~\cite{FMNR01} offer an efficient alternative to conventional 
highly-correlated {\em ab initio} methods to go beyond CASSCF and provide 
an accurate description of both dynamical and static electronic 
correlation. 
The key ingredient which determines the quality of a QMC calculation is the
many-body trial wave function which, in the present work,
is chosen of the Jastrow-Slater type with the particular form,
\begin{equation}
 \Psi^{\text{VMC}}_I = \Psi^{\text{CAS}}_I \prod_{A,i,j} {\cal J}(r_{ij}, r_{iA}, r_{jA}) \, ,
  \label{eq:QMC.2}
\end{equation}
where $r_{ij}$ denotes the distance between electrons $i$ and $j$, and
$r_{iA}$ the distance of electron $i$ from nucleus $A$. We use here
a Jastrow factor ${\cal J}$ which correlates pairs of electrons and
each electron separately with a nucleus, and employ different Jastrow
factors to describe the correlation with different atom types.
The determinantal component consists of a CAS expansion, $\Psi^{\text{CAS}}_I$,
and includes all possible CSFs obtained by placing 4 electrons in the
active space of 6 orbitals as explained above.

All parameters in both the Jastrow and the determinantal component of
the wave function are optimized by minimizing the energy.
Since the optimal orbitals and expansion coefficients in $\Psi^{\text{CAS}}_I$
may differ from the CASSCF values obtained in the absence of the Jastrow
factor ${\cal J}$, it is important to reoptimize  them in
the presence of the Jastrow component.

For a wave function corresponding to the lowest state of a given
symmetry, we follow the energy-minimization approach of
Ref.~\cite{UTFSH07}.
If the excited state is not the lowest in its symmetry, we obtain
the Jastrow and orbitals parameters which minimize the average energy
over the state of interest and the lower states, while the linear
coefficients in the CSF expansion ensure that orthogonality
is preserved among the states~\cite{F07}. Therefore, the wave functions
resulting from the state-average optimization will share the same
Jastrow parameters and the same set of orbitals but
have different linear coefficients. This scheme represents a
generalization of the approach of Ref.~\cite{SF04} where only the orbitals
were optimized and orthogonality was only approximately preserved.
The present approach is therefore superior to the one of Ref.~\cite{SF04}, which was
however already giving excellent results when tested on several singlet
states of ethylene and a series of prototypical photosensitive molecules~\cite{SF04,SBF04}.
We note that, when a CAS expansion is used in the absence of the Jastrow
component, the method is analogous to the CASSCF technique for the lowest
state of a given symmetry, and to a SA-CASSCF approach if the excited
state is not the lowest in its symmetry.

The trial wave function is then used in diffusion Monte Carlo (DMC), which produces
the best energy within the fixed-node approximation [i.e., the lowest-energy state
with the same zeros (nodes) as the trial wave function]. All QMC
results presented are from DMC calculations.


\section{Computational Details}
\label{sec:details}

\subsection{SCF and TD Details}

Calculations were carried out with two different computer programs, namely 
{\sc Gaussian} 03 \cite{Gaussian} and a development version of {\sc deMon2k} \cite{deMon2K}. 
We use the two programs in a complementary fashion as 
they differ in ways which are important for this work. In particular, {\sc Gaussian}
can carry out HF calculations while {\sc deMon2k} has no 
Hartree-Fock exchange. Similarly, {\sc Gaussian} can carry out time-dependent 
HF (TDHF) and configuration interaction singles (CIS) calculations 
which are not implemented in {\sc deMon2k}. In contrast, the Tamm-Dancoff 
approximation (TDA) for TDDFT may only be performed with {\sc deMon2k}.
The present version of {\sc deMon2k} is limited to 
the local density approximation for the exchange-correlation TDDFT kernel,
while the kernel is more general in {\sc Gaussian}. 
Finally, {\sc Gaussian} makes automatic use of symmetry and 
prints out the irreducible representation for each molecular orbital. 
The particular version of {\sc deMon2k} used in the present work does not have this 
feature so we rely on comparison with the output of {\sc Gaussian} calculations 
for this aspect of our analysis. 

All {\sc Gaussian} results reported here are calculated with the extensive 6-311++G**(2d,2p) 
basis set \cite{KBSP80,CCS83}. All calculations used default convergence criteria. 
The DFT calculations used the default grid for the exchange-correlation integrals.

The same basis set is used for {\sc deMon2k} calculations 
as in the Gaussian calculations. The GEN-A3* 
density-fitting basis set is used and density-fitting is carried out without 
imposing the charge conservation constraint. The SCF convergence cutoff  is set at 
$10^{-7}$. We always use the FIXED FINE option for the grid. 
Our implementation of TDDFT in {\sc deMon2k} is described in Ref.~\cite{IFP+05}.

We use two different density functionals, the local density 
approximation (LDA) in the parameterization of Vosko, Wilk, and Nusair \cite{VWN80} 
(referred to as SVWN5 in the {\sc Gaussian} input), and the popular B3LYP hybrid 
functional \cite{B3LYP}. The corresponding TDDFT calculations are referred to as TDLDA and 
TDB3LYP respectively. 

\subsection{CAS Details}
\label{CAS_Details}

All CASSCF calculations are performed with the program
{\sc GAMESS(US)}~\cite{Gamess}. In all SA-CASSCF calculations,
equal weights are employed for the two states.

We use scalar-relativistic energy-consistent Hartree-Fock
pseudopotentials~\cite{BFD07} where the carbon and oxygen 1$s$
electrons are replaced by a non-singular $s$-non-local pseudopotential
and the hydrogen potential is softened by removing the Coulomb
divergence.
We employ the Gaussian basis sets~\cite{BFD07} constructed
for these pseudopotentials and augment them with diffuse functions.
All calculations are performed with the cc-TZV contracted
(11$s$11$p$1$d$)/[3$s$3$p$1$d$] basis for carbon and oxygen,
augmented with two additional diffuse $s$ and $p$ functions with exponents
0.04402 and 0.03569 for carbon, and 0.07376 and 0.05974 for oxygen.
The $d$ polarization functions for carbon and oxygen are taken from
the cc-DZV set.
For hydrogen, the cc-DZV contracted (10$s$9$p$)/[2$s$1$p$] basis
is augmented with one $s$ diffuse function with exponent 0.02974.


\subsection{QMC Details}

The program package {\sc CHAMP}~\cite{Champ} is used for the
QMC calculations. We employ the same pseudopotentials and basis sets
as in the CASSCF calculations (see Sec.~\ref{CAS_Details}).

Different Jastrow factors are used to describe the correlation
with a hydrogen, an oxygen and a carbon atom. For each atom type, the
Jastrow factor consists of an exponential of the sum of two fifth-order
polynomials of the electron-nuclear and the electron-electron
distances, respectively~\cite{FU96}.
The parameters in the Jastrow factor and in
the determinantal component of the wave function are simultaneously
optimized by energy minimization following the scheme of Ref.~\cite{UTFSH07},
where we employ the simple choice $\xi=1$. For the
excited states with the same symmetry as the ground state, the ground-
and excited-state wave functions are optimized in a state-average manner
with equal weights for both states~\cite{F07}.
An imaginary time step of 0.075 H$^{-1}$ is used in the DMC calculations.

\section{Results}
\label{sec:results}

\begin{figure}
 \includegraphics[angle=0,scale=0.45]{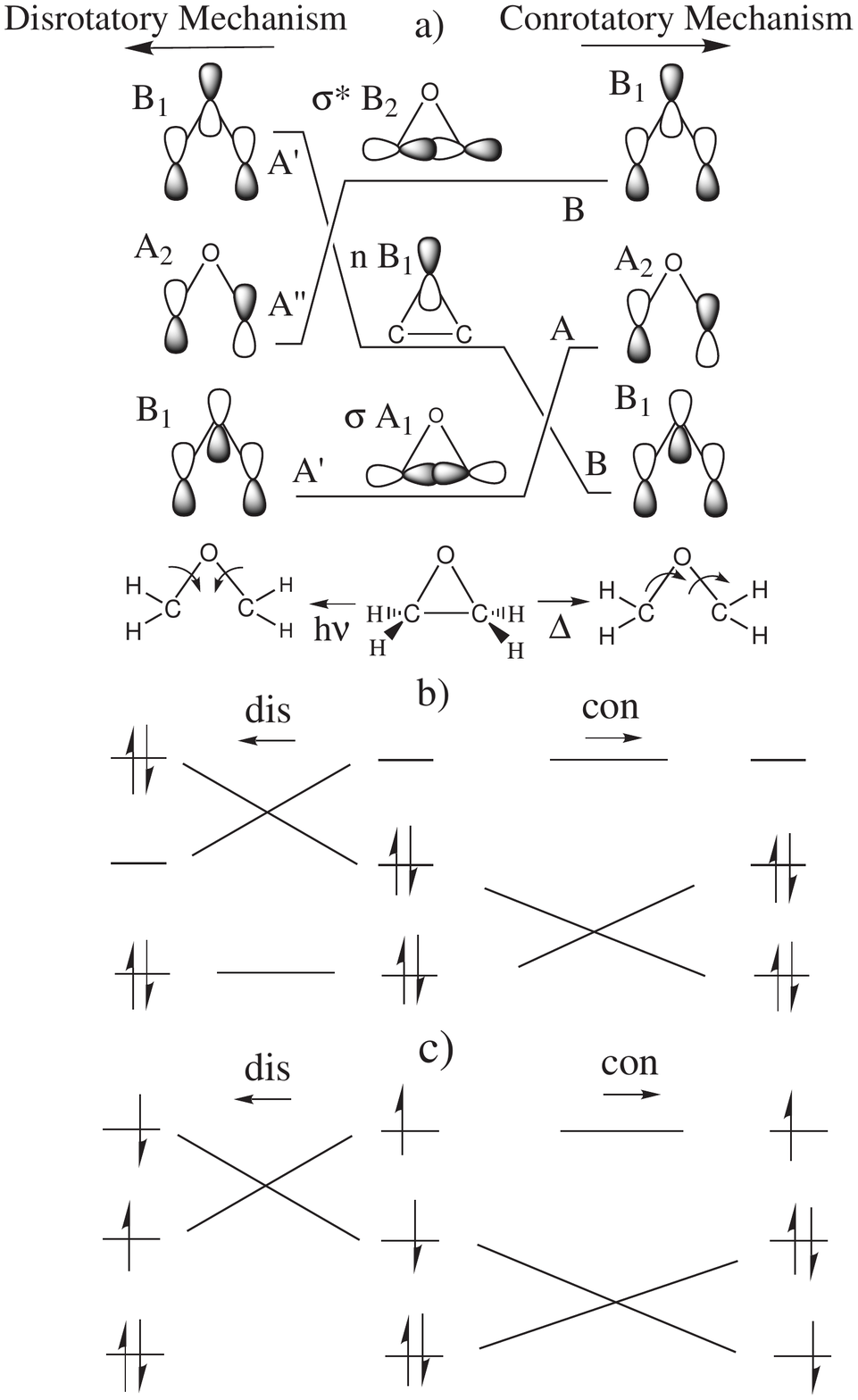}
 \caption{a) Woodward-Hoffmann orbital correlation scheme.  
 Symmetry labels are for the $C_{2v}$ point group in the case of reactants and products, 
 for the $C_s$ point group along the disrotatory pathway, and for the $C_2$ point group 
 along the conrotatory pathway.  b) Thermal ring opening. c) Photochemical ring opening.
 Symmetry labels depend upon the labels of the $x$, $y$, and $z$ coordinates. In this article,
 the COC ring lies in the  $(y, z)$-plane and the $z$-axis coincides with the $C_2$ axis,
 in agreement with the IUPAC convention\cite{M55}.
 }
 \label{fig:WHcorrdiag}
\end{figure}

The Woodward-Hoffmann (WH) rules for orbital control of symmetry in electrocyclic reactions 
were an important motivation 
in the 1970s and 1980s to seek stereospecific photochemical ring-opening reactions of 
oxiranes \cite{H71b,GP76,L76,H77,HRS+80,P87}.  
Since some oxiranes
(notably diphenyl oxirane) do appear to follow the WH rules for thermal and photochemical
ring opening, the WH rules might seem like the obvious place to begin our study of 
symmetric ring opening pathways in oxirane.  This is somewhat counterbalanced by
the fact that oxirane itself is an exception to the WH rules for photochemical
ring opening and by the fact that it is now clear that the WH rules do not apply 
nearly as well to photochemical as to thermal reactions 
(Appendix~\ref{sec:photochem}).   Nevertheless, we will take the WH model as a first
approximation for understanding and begin by describing the model for the particular
case of oxirane.

Many models at the time that Woodward and Hoffmann developed their theory \cite{WH65a,WH65b,WH69} 
were based upon simple H\"uckel-like $\pi$-electron models, so, not surprisingly, the 
historical WH model for oxirane uses a three-orbital model consisting of one $p$-orbital 
on each of the oxygen and two carbon atoms.  (See Fig.~\ref{fig:WHcorrdiag}.)  Elementary 
chemical reasoning predicts that the closed cycle should form three molecular orbitals, 
$\sigma$, $n$, $\sigma^*$, in increasing order of energy.  Similarly the open structure 
has the ``particle-in-a-box'' orbitals familiar from simple H\"uckel theory.  Woodward and 
Hoffmann observed that a reflection plane ($\sigma$) of symmetry is preserved along 
the disrotatory reaction pathway while a $C_2$ rotation symmetry element is preserved 
along the conrotatory reaction pathway.  This observation allows the reactant-product 
molecular orbital correlation diagram to be completed by using the fact that the
symmetry representation of each orbital is preserved within the relevant symmetry group 
of the molecule along each reaction path (WH principle of conservation of orbital symmetry) 
and connecting lowest orbitals with lowest orbitals.  Figure~\ref{fig:WHcorrdiag} shows 
that a conrotatory ({\em con}) thermal reaction connects ground-state configurations in the
reactant and product while a disrotatory ({\em dis}) thermal reaction connects the reactant 
ground-state configuration with an electronically excited-state configuration.  Thus, the 
{\em con} thermal reaction is expected to be prefered over the {\em dis} reaction.  
Also shown in Fig.~\ref{fig:WHcorrdiag} is the photochemical reaction beginning with the 
$n \rightarrow \sigma^*$ excited state.  In this case, the {\em dis} mechanism is expected 
to be prefered since the {\em con} mechanism leads to a still higher level of excitation 
in the product than in the reactant molecule while the level of excitation is preserved 
along the {\em dis} pathway.

Let us now see whether this orbital model is reflected first in the vertical
absorption spectrum of oxirane, then in the $C_{2v}$ ring opening pathway, and finally
in the {\em con} and {\em dis} ring-opening pathways.  At each step, the performance of
TDDFT will be assessed against experiment and against other levels of theory.

\subsection{Absorption Spectrum}

\begin{figure}
  \includegraphics[angle=0,scale=0.50]{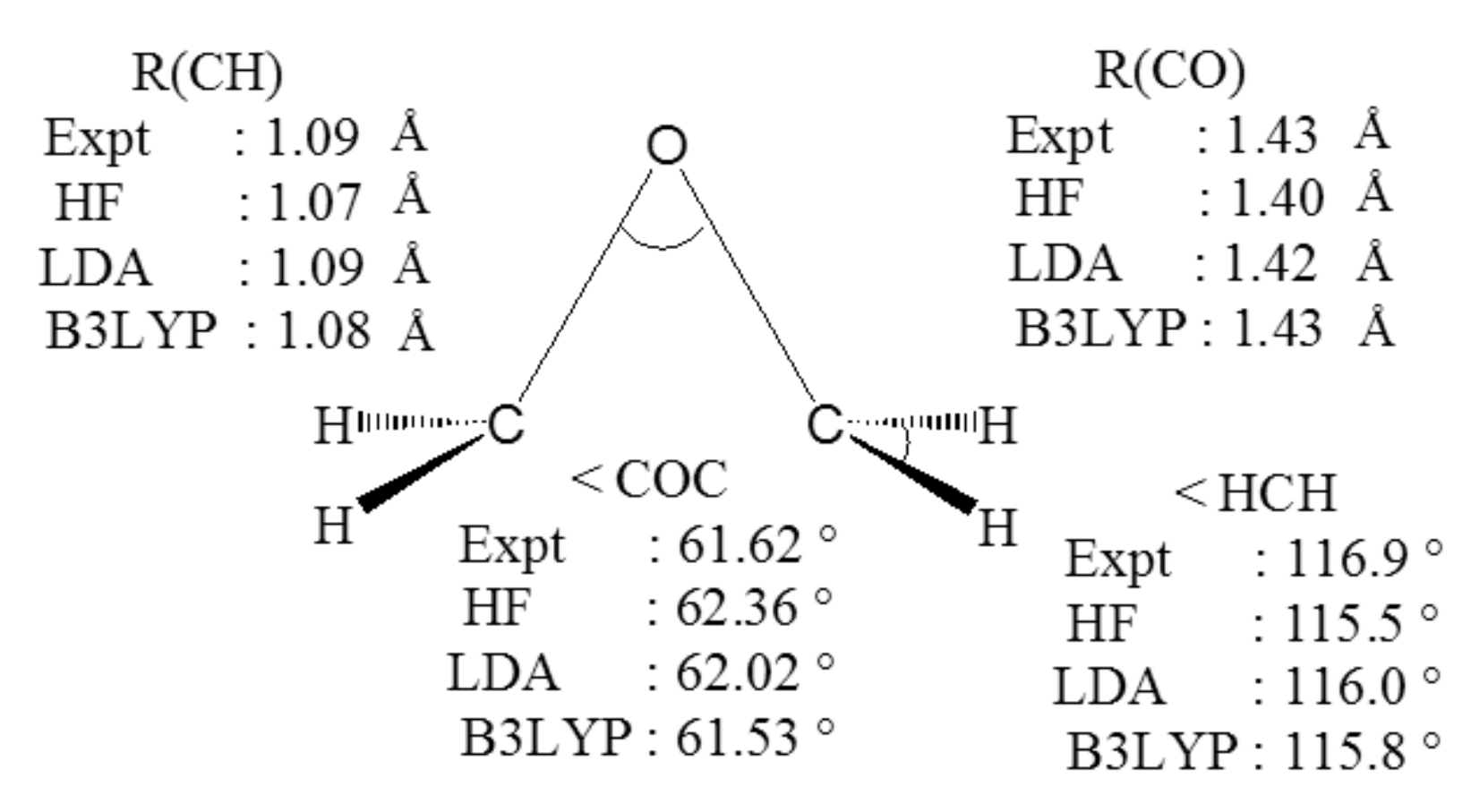}
 \caption{Comparison of HF/6-311G$^{**}$(2d,2p), B3LYP/6-311G$^{**}$(2d,2p), 
 and LDA/6-311G$^{**}$(2d,2p) optimized geometries with the experimental 
 gas phase geometry from Ref.~\cite{H47}.  Note that the structure has $C_{2v}$ symmetry.}
\label{fig:optgeom}
\end{figure}

The first step towards calculating the electronic absorption spectrum of oxirane is the
optimization of the geometry of the gas phase molecule.  This was carried out using the HF
method and DFT using the LDA and B3LYP functionals.  The calculated results are compared 
in Fig.~\ref{fig:optgeom} with the known experimental values.  
It is seen that electron correlation included in DFT
shortens bond lengths, bringing the DFT optimized geometries into considerably better 
agreement with the experimental geometries than are the HF optimized geometries.  This better
agreement between DFT and experiment also holds for bond angles.  There is not
much difference between the LDA and B3LYP geometries with the exception of the COC
bond angle which is somewhat better described with the B3LYP than with the LDA
functional. 


\begin{table}
\caption{
Principal oxirane singlet excitation energies (eV) and oscillator strengths (unitless).
}
\label{tab:spect}
\begin{tabular}{cccc}
\hline \hline
 TDHF         &  TDLDA         &  TDB3LYP & Expt. \\ 
\hline
9.14 (0.0007) &  6.01 (0.0309) &  6.69 (0.0266) & 7.24(s)
                                              \footnotemark[1] \footnotemark[2] \footnotemark[3] \\
9.26 (0.0050) &  6.73 (0.0048) &  7.14 (0.0060) & 7.45(w)\footnotemark[2]  \\
9.36 (0.0635) &  6.78 (0.0252) &  7.36 (0.0218) & 7.88(s)\footnotemark[1], 7.89(s) \footnotemark[2] \\
9.56 (0.0635) &  7.61 (0.0035) &  7.85 (0.0052) &            \\
9.90 (0.0478) &  7.78 (0.0304) &  8.37 (0.0505) &            \\
9.93 (0.0935) &  8.13 (0.0014) &  8.39 (0.0168) &            \\
8.15 (0.0405) &  8.40 (0.0419) &                &            \\
\hline \hline
\end{tabular}
\footnotetext[1]{Gas phase UV absorption spectrum \cite{LD49}.}
\footnotetext[2]{Obtained by a photoelectric technique \cite{LW58}.}
\footnotetext[2]{Gas phase UV absorption spectrum \cite{FAHP59}.}
\end{table}


Several experimental studies of the absorption spectrum of oxirane are available~\cite{LD49,LW58,FAHP59,BRK+69,BBG+92}.
The positions of the principal electronic excitations were identified early on but their 
assignment took a bit longer. The accepted interpretation is based on the study by
Basch {\em et al}.~\cite{BRK+69} who combined information from vacuum ultraviolet spectra,
photoelectron spectra, and quantum chemical computations, and assigned the observed transitions to
O($n$) $\rightarrow$ $3s$ and $3p$ Rydberg transitions.  Let us see how this
is reflected in our calculations.

While theoretical absorption spectra are often shifted with respect to
experimental absorption spectra, we expect to find two strong absorptions in the low energy
spectrum, separated from each other by about 0.65 eV.  Vertical absorption spectra are
calculated using TDHF, TDLDA, and TDB3LYP, all at the B3LYP-optimized geometry.  
The results are shown in Table~\ref{tab:spect}.  TDB3LYP is the method in best agreement 
with experiments, yielding two strong absorptions red-shifted 
from experiment by about 0.5 eV and separated by 0.67 eV.  The TDLDA method is qualitatively similar, apart
from a stronger red shift.  In particular, the TDLDA method shows two strong absorptions red-shifted
from experiment by about 1.1 eV and separated by 0.78 eV.  In contrast, the TDHF spectrum is
blue-shifted by about 2 eV and is otherwise of questionable value for interpreting the 
experimental spectrum.  In what follows, we have decided to assign the lowest three absorptions
using the results of our TDB3LYP calculations.

\begin{figure}
  \includegraphics[angle=0,scale=0.6]{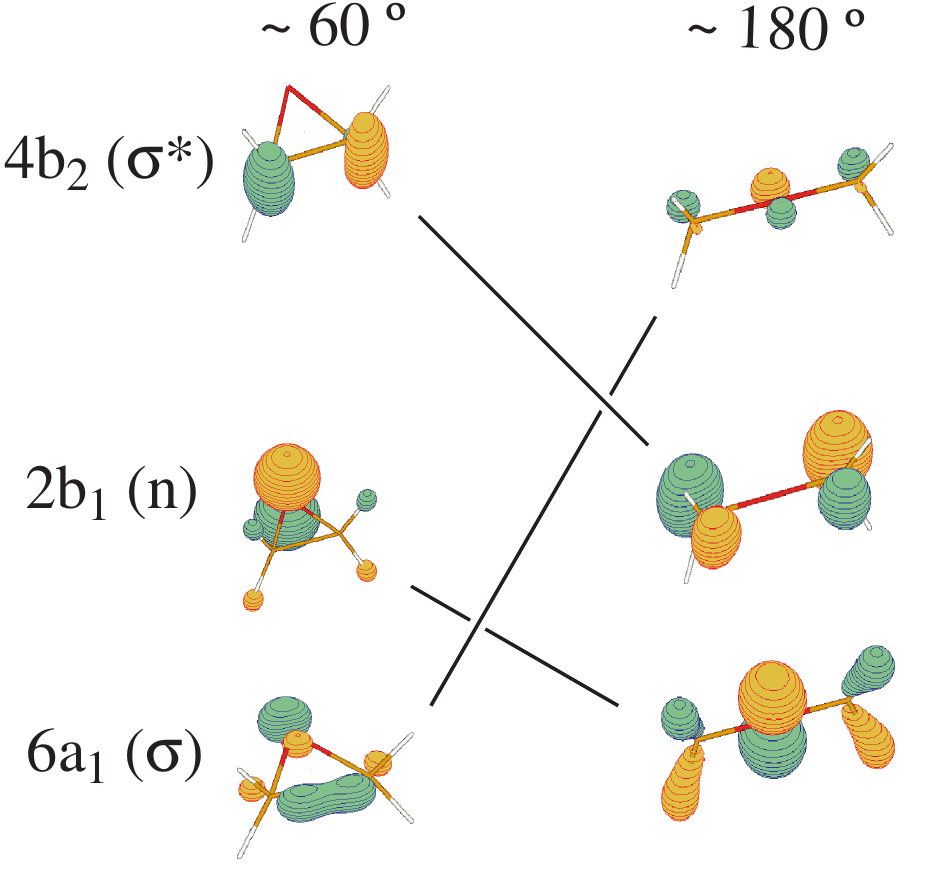}
  \caption{B3LYP MOs. Left: ring structure.  Right: open structure.}
  \label{fig:orbitals}
\end{figure}

This first involves an examination of the B3LYP molecular orbitals (MOs).  These orbitals are 
shown in Fig.~\ref{fig:orbitals}.  The electronic configuration of the ring structure is,
\begin{equation}
   \cdots [6a_1(\sigma)]^2 [2b_1(n)]^2 [7a_1(3s)]^0 [4b_2(\sigma^*)]^0 \cdots \, .
   \label{eq:results.1}
\end{equation}
The group theoretic MO labels ($a_1$, $a_2$, $b_1$, and $b_2$) correspond to representations
of the $C_{2v}$ symmetry group.  The additional labels, $\sigma$, $n$, and $\sigma^*$, show our
chemical interpretation of the B3LYP orbitals and their correspondence with the MOs in the 
WH three-orbital model.  

\begin{figure}
  \includegraphics[angle=0,scale=0.50]{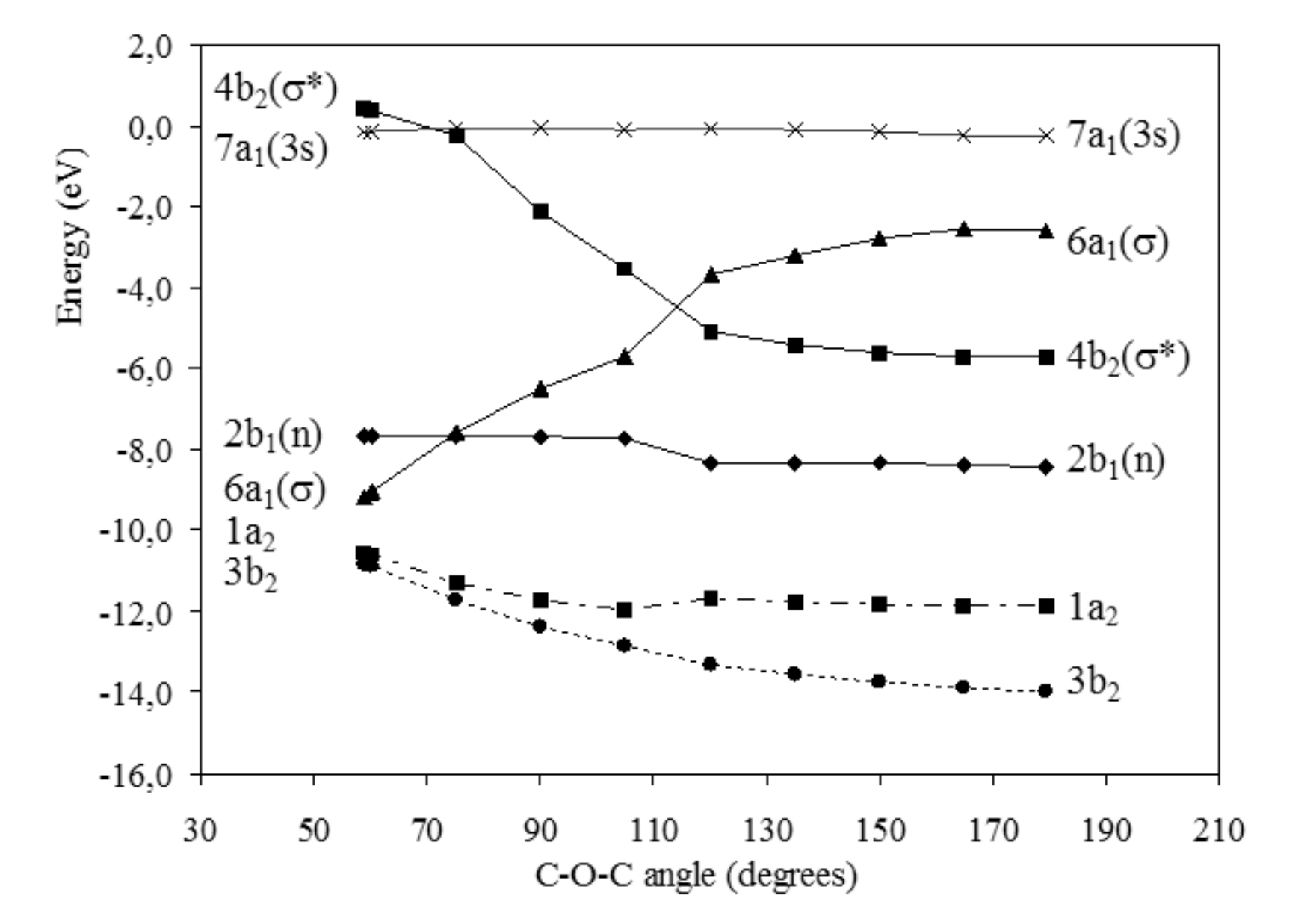}
  \caption{Walsh diagram for $C_{2v}$ ring opening calculated at the B3LYP level. To construct 
  this diagram, the COC bond angle was varied and all other geometric parameters were relaxed 
  within the constraint of $C_{2v}$ symmetry.  
  The HOMO is the $2b_1$ orbital on the left hand side and the $4b_2$ orbital on the right hand side.
  }
  \label{fig:Walsh_C2v}
\end{figure}

\begin{figure}[t]
  \includegraphics[angle=0,scale=0.45]{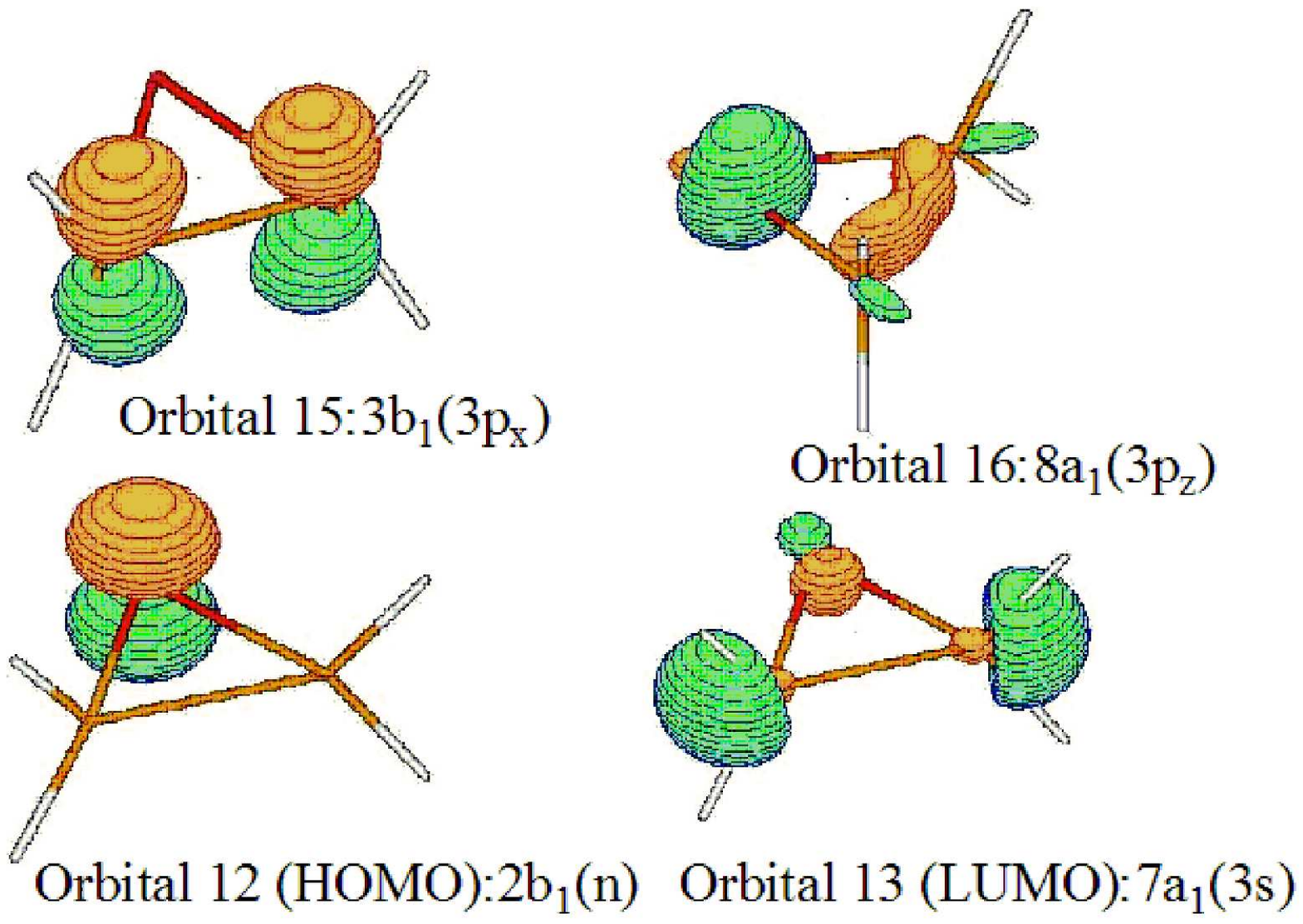}
  \caption{B3LYP MOs implicated in the principal UV absorptions.}
  \label{fig:excite_orbs}
\end{figure}

Confirmation of the chemical nature of our B3LYP MOs was obtained by constructing a Walsh
diagram for $C_{2v}$ ring opening.  This graph of MO energies as a function of ring-opening angle
($\alpha$) is shown in Fig.~\ref{fig:Walsh_C2v}.  As the ring opens, the C($2p$)
$6a_1(\sigma)$ bond breaks and so increases in energy.  At the same
time, the C($2p$) $4b_2(\sigma^*)$ antibond becomes less antibonding and so decreases in energy.
The O($2p$) lone pair $2b_1(n)$ is not involved in bonding and so maintains a roughly constant energy
throughout the ring-opening process.  Experimental information about occupied MOs is 
available from electron momentum spectroscopy (EMS) via the target Kohn-Sham 
approximation \cite{DCCS94}.  The ordering of the B3LYP occupied orbitals is consistent with the 
results of a recent EMS study of oxirane \cite{WMWB99}.  In particular, the two highest 
energy occupied orbitals are seen to have a dominant $p$-type character.  The interpretation 
of the unoccupied orbitals is more problematic in that the B3LYP unoccupied orbitals in our
calculations are not bound (i.e., have positive orbital energies).  We are thus attempting to
describe a continuum with a finite basis set.  It is thus far from obvious that the unoccupied 
orbital energies will converge to anything meaningful as the finite basis set becomes increasingly
complete.  
The Walsh diagram shows that the $4b_2(\sigma^*)$ unoccupied orbital becomes
bound for ring opening angles beyond about 80$^\circ$.  It is also rather localized and hence
it makes sense to assign it some physical meaning.  This is certainly consistent with previous
HF studies using the STO-3G minimal basis set \cite{BSD79a} and the more extensive 6-31G**
basis sets \cite{HSS01}.  Since our 6-311++G**(2d,2p) basis set is even larger, it is not too
surprising that we find an additional unoccupied orbital, namely the $7a_1(3s)$ orbital shown
in Fig.~\ref{fig:excite_orbs}.  Although apparently at least partially localized, this orbital 
remains unbound at all bond angles in the Walsh diagram and care should be taken not to 
overinterpret its physical nature.  Nevertheless this $7a_1(3s)$ orbital intervenes in an important 
way in the interpretation of our calculated electronic absorption spectra and it is upon
analysis of the spectra that we will be able to associate this orbital with the $3s$ Rydberg
state.

One should also be conscious in using the calculated TDB3LYP absorption spectrum to assign the
experimental gas phase UV spectrum that the TDDFT ionization continuum begins at minus the 
value of the HOMO orbital energy \cite{CJCS98}.  The value of $-\epsilon_{\text{HOMO}}$ obtained
in the different calculations is 6.40 eV with the LDA, 7.68 eV with the B3LYP hybrid functional, 
and 12.27 eV with HF.  As expected from Koopmans' theorem, the HF value of $-\epsilon_{\text{HOMO}}$
is in reasonable agreement with the experimental ionization potential of 10.57 eV \cite{NCK77}.
The presence of a fraction of HF exchange in the B3LYP hybrid functional helps to explain why
its value of $-\epsilon_{\text{HOMO}}$ lies between that of the pure DFT LDA and that of  HF.
As far as our TDB3LYP calculations are concerned, the value of $-\epsilon_{\text{HOMO}}$
means that assignment of experimental excitation energies higher than 7.7 eV should be avoided.
Fortunately this still allows us to assign the first singlet excitation energies.  Examination 
of the TDB3LYP coefficients yields the following assignments for the three principal UV absorption peaks:
\begin{eqnarray}
  \mbox{6.69 eV} & : &  1^1B_1[2b_1(n) \rightarrow 7a_1(3s)] \\
  \mbox{7.14 eV} & : &  2^1B_1[2b_1(n) \rightarrow 8a_1(3p_z)] \\
  \mbox{7.36 eV} & : &  2^1A_1[2b_1(n) \rightarrow 3b_1(3p_x)] \, .
  \label{eq:results.2}
\end{eqnarray}
Comparison with the experimental assignment of Basch {\em et al}.\ justifies the identification
of these orbitals with the Rydberg orbitals $3s$, $3p_z$, and $3p_x$.

It is worth pointing out that the expected $^1[2b_1(n),4b_2(\sigma^*)]$ excitation is
of $^1A_2$ symmetry and as such corresponds to a spectroscopically forbidden transition.
It is in any event fairly high in energy (9.77 eV with TDHF).  The $^1[6a_1(\sigma),4b_2(\sigma^*)]$
transition has $^1B_2$ symmetry and so is spectroscopically allowed, but is still found at
fairly high energy in our calculations (8.15 eV with TDLDA, 8.37 with TDB3LYP, and 9.93 with
TDHF).


\begin{figure*}[tb]
    \includegraphics[width=\columnwidth]{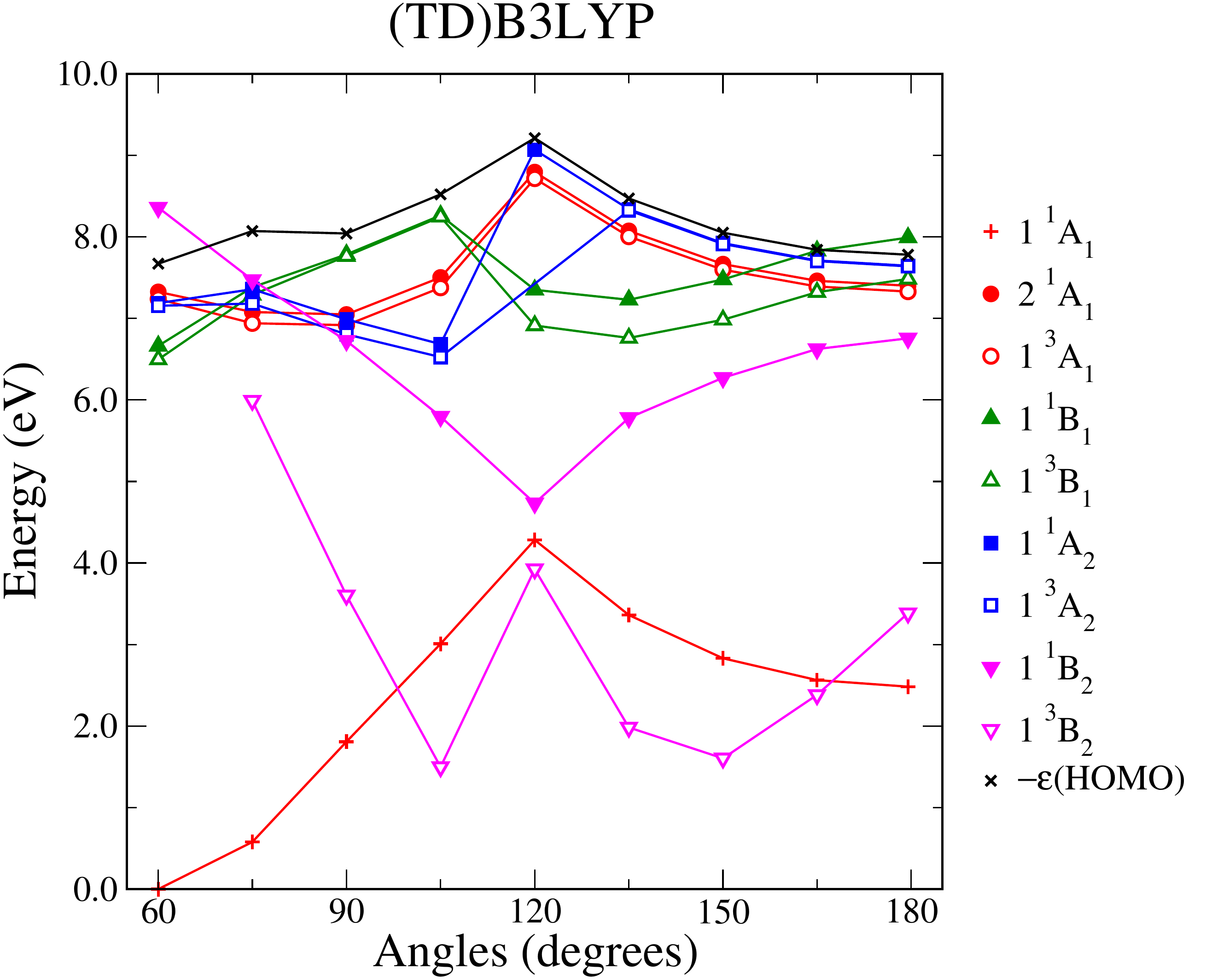}
    \includegraphics[width=\columnwidth]{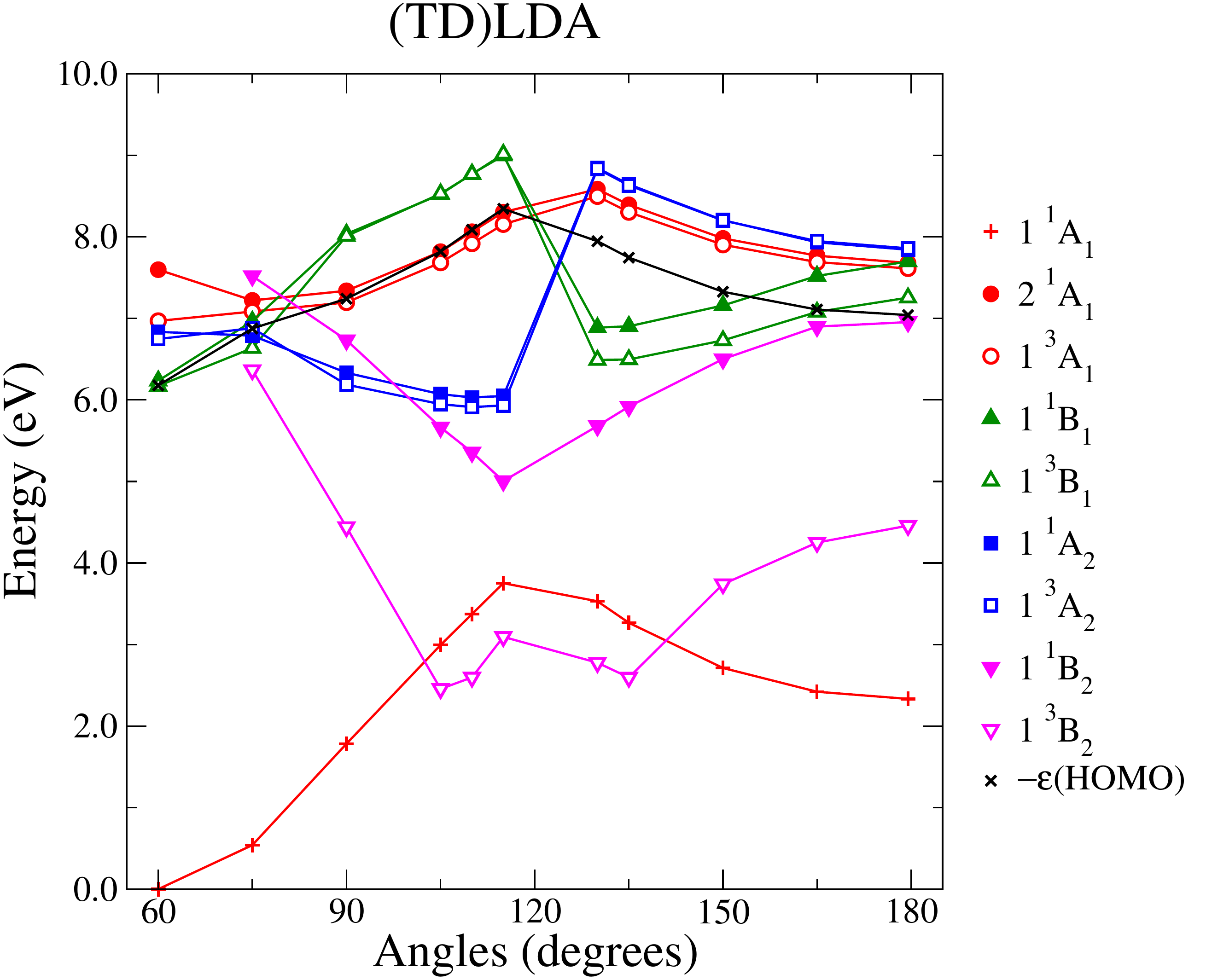}
   \caption{$C_{2v}$ ring opening curves: ground state ($1 ^1A_1$) curve 
   calculated using the B3LYP ({\sc Gaussian}) or the LDA ({\sc deMon2k}) functional, 
   lowest excited state curve of each symmetry ($2 ^1A_1$,
   $1 ^3A_1$, $1 ^1B_1$, $1 ^3B_1$, $1 ^1A_2$, $1 ^3A_2$, $1 ^1B_2$, and $1 ^3B_2$) calculated using
   the TDB3LYP ({\sc Gaussian}) and the TDLDA ({\sc deMon2k}) excitation energies added to the B3LYP and
   the LDA ground state energy, respectively.   
   The energy zero has been chosen to be the ground state energy for the 60$^\circ$ structure.
   Note that the ``negative excitation energies'' for the $1 ^3B_2$ state relative to the 
   ground state are really imaginary excitation energies (see text).  Also
   shown is the TDDFT ionization threshold at $-\epsilon_{\text{HOMO}}$.}
   \label{fig:all_TDDFT_curves}
\end{figure*}

\subsection{$C_{2v}$ Ring-Opening}

We now consider how DFT performs for describing the ground state and how well TDDFT
performs for describing the lowest excited state of each symmetry for $C_{2v}$ ring-opening 
of oxirane.  Comparisons are made against CASSCF and against high-quality DMC energies
calculated at the same geometries.  All geometries along the $C_{2v}$ pathway have been fully
optimized at each O~-~C~-~O ring-opening angle ($\alpha$) using the B3LYP functional.
The orbital energies as a function of ring opening angle have already been given in the
Walsh diagram (Fig.~\ref{fig:Walsh_C2v}).

Fig.~\ref{fig:all_TDDFT_curves} shows the TDB3LYP and TDLDA curves
for the ground ($1 ^1A_1$) state and the lowest excited-states of each symmetry 
($2 ^1A_1$, $1 ^3A_1$, $1 ^1B_1$, $1 ^3B_1$, $1 ^1A_2$, $1 ^3A_2$, $1 ^1B_2$, and $1 ^3B_2$). 
Several things are worth noting here.  The {\em first} point is that, some of the differences 
between the TDB3LYP and TDLDA curves are apparent, not real, as
the points for the two graphs were not calculated at exactly the same angles. In
particular, the TDLDA misses the point at 120$^\circ$ 
where we encountered serious convergence difficulties due to a 
quasidegeneracy of $\sigma$ and $\sigma^*$ orbitals (Fig.~\ref{fig:Walsh_C2v}).  Under 
these circumstances, the HOMO, which suffers from self-interaction errors, can lie higher
than the self-interaction-free LUMO.  As the program tries to fill the orbitals according
to the usual {\em aufbau} principle, electron density sloshes on each iteration between the two
orbitals making convergence impossible without special algorithms.  The TDB3LYP calculations
were found to be easier to converge, presumably because they have less self-interaction error.

The {\em second} point is that only {\sc deMon2k} TDLDA calculations are reported 
here, although we have also calculated TDLDA curves using 
{\sc Gaussian} and found similar results when both programs printed out the same
information.  Unfortunately, {\sc Gaussian} did not
always print out the lowest triplet excitation energy, so we prefer to report our more
complete {\sc deMon2k} results.  

The {\em third} point is the presence of a triplet instability.  This means 
that, going away from the equilibrium geometry, the {\em square} of the first triplet excitation 
energy decreases, becomes zero, and then negative.  The exact meaning of 
triplet instabilities will be discussed further below.  For now, note that
{\em we follow the usual practice for response calculations by indicating an imaginary excitation 
energy as a negative excitation energy.}
However it is important to keep in mind that a negative excitation energy in this context
is only a common convention and not a physical reality.  
Incidentally, we presume that triplet instability is the source of the difficulty with the 
{\sc Gaussian} output mentioned above--the associated coefficients become more complicated 
and {\sc Gaussian} may have difficulty analyzing them.  

A {\em fourth} and final point 
is that the TDDFT ionization threshold occurs at minus the value of the HOMO
energy which is significantly underestimated with ordinary functionals such as the LDA and
B3LYP \cite{CJCS98}.  This artificially low ionization threshold is indicated in the two graphs
in Fig.~\ref{fig:all_TDDFT_curves}.  In the TDB3LYP case, the TDDFT ionization threshold is high
enough that it is not a particular worry.  In the TDLDA case, the TDDFT ionization threshold is
lower by about 1 eV.  This may explain some of the quantitative differences between the high-lying
TDB3LYP and TDLDA curves, although by and large the two calculations give results in reasonable
qualitative, and even semi-quantitative, agreement.

We now wish to interpret the TDDFT curves.  TDDFT excitations may be characterized 
in terms of single electron excitations from occupied to unoccupied MOs \cite{C95}.
Unlike in HF, unoccupied and occupied orbitals in pure DFT (e.g., the LDA) 
see very similar potentials and hence the same number of electrons.  This means that the
orbitals are preprepared for describing electron excitations and that simple orbital
energy differences are often a good first approximation to describing excitation energies.
In the two-orbital model and pure DFT, the singlet, $\omega_S$, and triplet, $\omega_T$, 
TDDFT excitation energies for the transition from orbital $i$ to orbital $a$ are,
\begin{eqnarray}
   \omega_S & = & \sqrt{\left(\epsilon_a - \epsilon_i\right)
   \left[\left(\epsilon_a - \epsilon_i\right) + 2(ia \vert 2f_H + f_{xc}^{\uparrow,\uparrow}
   + f_{xc}^{\uparrow,\downarrow} \vert ai ) \right] } \nonumber \\
   \omega_T & = & \sqrt{\left(\epsilon_a - \epsilon_i\right)
   \left[\left(\epsilon_a - \epsilon_i\right) + 2(ia \vert f_{xc}^{\uparrow,\uparrow}
   - f_{xc}^{\uparrow,\downarrow} \vert ai ) \right] } \, . 
   \label{eq:results.3}
\end{eqnarray}
In the TDA, this becomes
\begin{eqnarray}
   \left(\omega_S\right)_{\text{TDA}} & = & \left(\epsilon_a - \epsilon_i\right)
    + (ia \vert 2f_H + f_{xc}^{\uparrow,\uparrow} + f_{xc}^{\uparrow,\downarrow} \vert ai ) 
    \nonumber \\
   \left(\omega_T\right)_{\text{TDA}} & = & \left(\epsilon_a - \epsilon_i\right)
    + (ia \vert f_{xc}^{\uparrow,\uparrow} - f_{xc}^{\uparrow,\downarrow} \vert ai )  \, , 
   \label{eq:results.4}
\end{eqnarray}
from which it is clear from the sizes and signs of the integrals that,
\begin{equation}
   \left( \omega_T \right)_{\text{TDA}} \leq \epsilon_a - 
    \epsilon_i \leq \left( \omega_S \right)_{\text{TDA}}\, .
   \label{eq:results.5}
\end{equation}
For Rydberg states the inequalities become near equalities because the overlap of orbitals
$i$ and $a$ goes to zero.
While no longer strictly valid, these general ideas are still good starting points for
understanding results obtained with the hybrid functional B3LYP.

\begin{figure*}[htb]
   \includegraphics[width=\columnwidth]{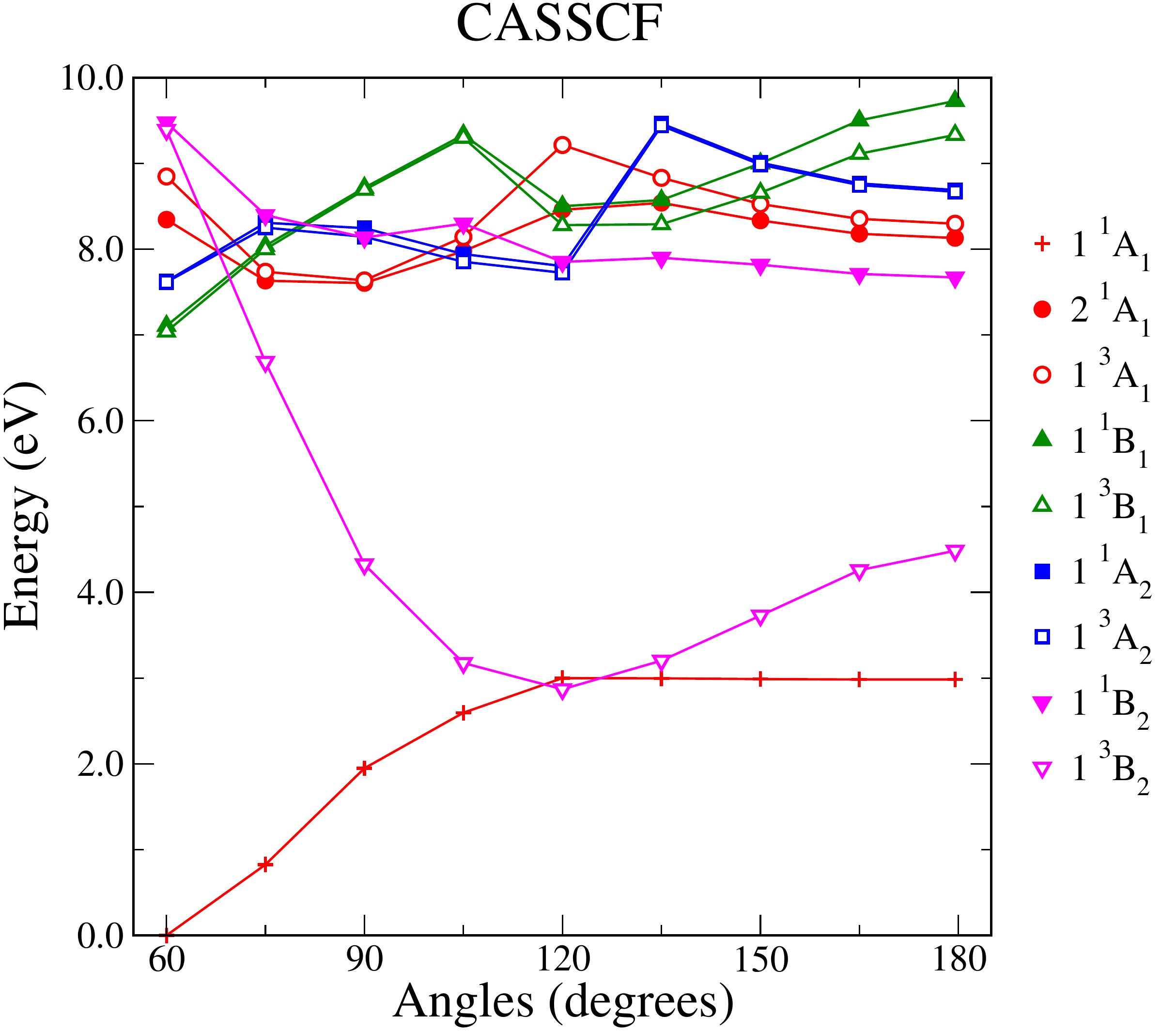}
   \includegraphics[width=\columnwidth]{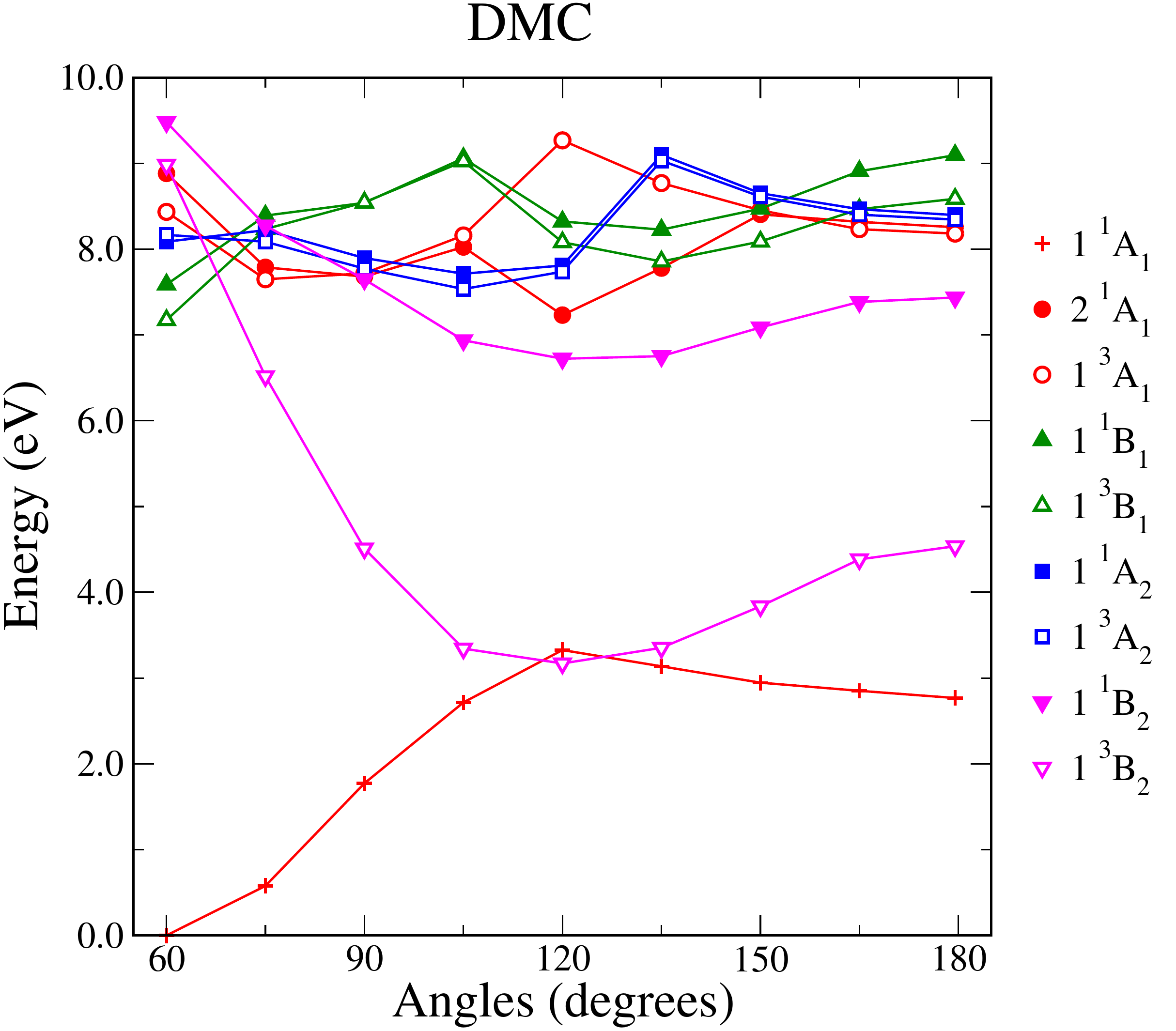}
   \caption{$C_{2v}$ ring opening curves calculated with CASSCF and DMC: curves for the lowest state 
   of each symmetry ($1 ^1A_1$, $1 ^3A_1$, $1 ^1B_1$, $1 ^3B_1$, $1 ^1A_2$, $1 ^3A_2$, $1 ^1B_2$, 
   and $1 ^3B_2$) are calculated using CAS(4,6) without state averaging, while the $2 ^1A_1$ is
   the result of adding the excitation energy from a state averaged calculation to the ground state
   $1 ^1A_1$ curve calculated without state averaging.  The energy zero has been 
   chosen to be the ground state energy for the 60$^\circ$ structure.  Note that the 
   negative excitation energies for the $1 ^3B_2$ state relative to the 
   ground state are really negative excitation energies. 
   Numerical DMC energies are listed in Appendix~\ref{sec:suppl}.
  }
  \label{fig:CAS_and_DMC_curves}
\end{figure*}

\begin{figure}
   \includegraphics[width=\columnwidth]{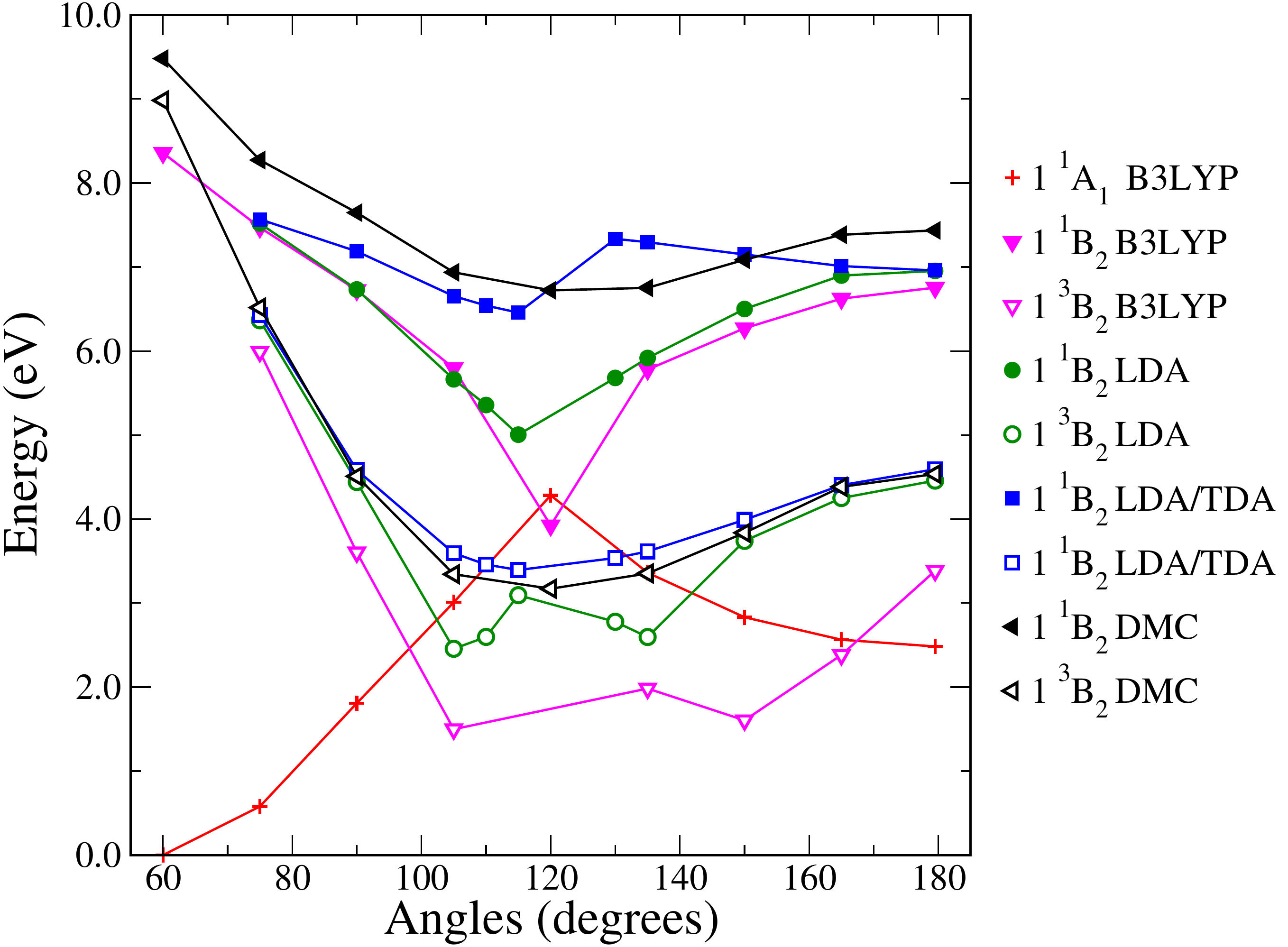}
   \caption{$C_{2v}$ ring opening curves: $1^3B_2$ and $1^1B_2$ states. Note that the ground state
   ($1^1A_1$) curve has only been shown for the B3LYP calculation since the LDA
   curve is nearly identical.}
  \label{fig:B2_curves}
\end{figure}

The frontier MOs shown in the Walsh diagram (Fig.~\ref{fig:Walsh_C2v}) supplemented with the 
addition of the two additional Rydberg orbitals, $3b_1(3p_x)$ and $5b_2(3p_y)$, provides a useful 
model not only for interpreting the results of our TDDFT calculations, but also for 
constructing the active space necessary for the CASSCF calculations used to construct the 
QMC wavefunctions.   
Fig.~\ref{fig:CAS_and_DMC_curves}
shows the results of our best QMC calculation and of the results of the CAS(4,6) calculation
on which it is based.  For the most part, the CAS and DMC curves differ quantitatively but 
not qualitatively.  The exceptions are the $1 ^3A_1$ and $1 ^1B_2$ curves where
the DMC results, which are significantly lower than the corresponding CAS(4,6) results at some
geometries, contain important amounts of dynamic correlation not present at the
CAS(4,6) level of calculation.  In the remaining graphs (except for Fig.~\ref{fig:B2_curves}) 
we will suppress CAS(4,6) curves in favor of presenting only the higher quality DMC 
curves since these offer the better comparison with TDDFT.

We now give a detailed comparison of our TDDFT results with
DMC methods, dividing the states into three sets.
The first set consists of states where double excitations 
(lacking in the TDDFT adiabatic approximation) are likely to be important.  
The second set consists of states which show the effects of triplet instabilities.
And the third set consists of Rydberg excitations.

The states of $A_1$ symmetry are those most likely to be affected by the absence of double
excitations.  
Although not shown here, we have constructed the two HF curves obtained by 
enforcing a double occupancy of either the $6a_1(\sigma)$ or the
$4b_2(\sigma^*)$ orbital.  Both single determinants have $^1A_1$ symmetry. 
The two curves thus generated simply cross at about 120$^\circ$.
In the absence of configuration mixing, the ground state curve always follows
the lower state and shows an important cusp at 120$^\circ$.  Introducing configuration
mixing via a CASSCF calculation leads to an avoided crossing.

\begin{figure}[bht]
    \includegraphics[width=\columnwidth]{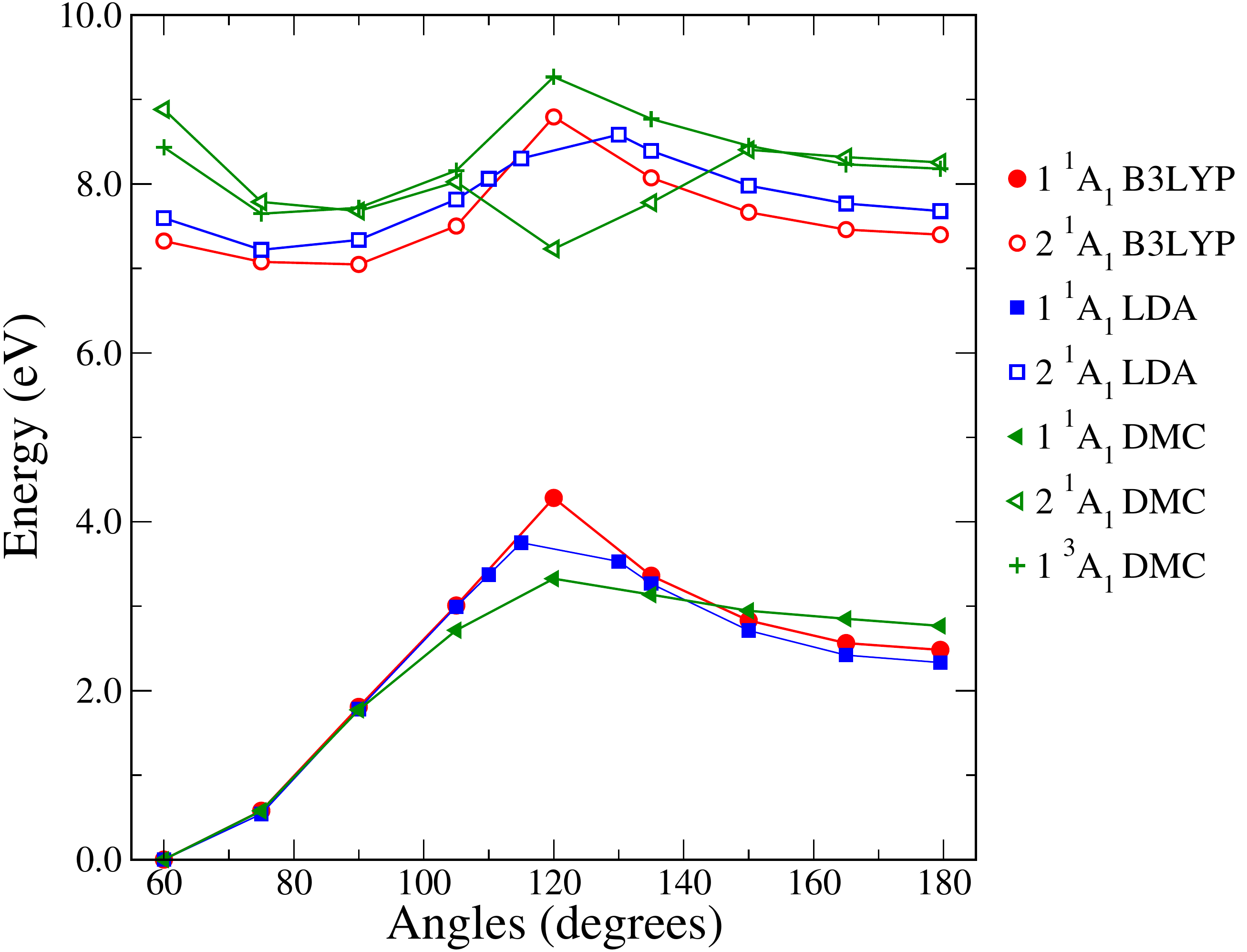}
  \caption{$C_{2v}$ ring opening curves: $1^1A_1$ and $2^1A_1$ states. Note that the $1^3A_1$ state
   has also been included at the DMC level of calculation.}
  \label{fig:1A1_curves}
\end{figure}
Now, DFT is different from HF because DFT is exact when the functional $E_{xc}$ is exact while
HF always remains an approximation.  On the other hand, the Kohn-Sham equations of modern DFT
resemble the HF equations and tend to inherit some of their faults when approximate functionals
are used.  Fig.~\ref{fig:1A1_curves} shows the comparable curves at the TDDFT and DMC levels.
As expected, the unphysical cusp is absent in the $1 ^1A_1$ ground state curve at the DMC 
level of calculation.  The unphysical cusp is present at both the B3LYP and LDA levels where
significant divergences between the DFT and DMC calculations occur between about 100$^\circ$
and 150$^\circ$.  However, this region is very much reduced compared to what we have observed to
happen for the HF curve where significant divergences beginning at about 75$^\circ$.  
This is consistent with the idea that even DFT with approximate functionals
still includes a large degree of electron correlation.

As to the $2 ^1A_1$ excited state curve, only the DMC calculation shows an indication of 
an avoided crossing.  The inclusion of the $1^3A_1$ excited state curve calculated
at the DMC level helps to give a more complete understanding of the curve for this state.  
Below about 100$^\circ$ and above about 150$^\circ$, the $2 ^1A_1$ and $1^3A_1$ states 
have nearly the same energy, consistent with the idea that these correspond to the 
$^1[a_1(\sigma),a_1(3s)]$ Rydberg transition below 110$^\circ$ and to the 
$^1[b_2(\sigma^*),b_2(3p_z)]$ Rydberg transition above 150$^\circ$.  
In between the $^1[a_1(\sigma)^2,b_2(\sigma^*)^2]$ two-electron valence excitation 
cuts across the $^1A_1$ manifold. The TDDFT calculations are qualitatively
capable of describing these one-electron Rydberg excitations but not of describing the 
two-electron excitation in the bond-breaking region.  As expected, the cusp in the
DFT $1^1A_1$ ground state curve is simply reflected as cusps in the TDDFT $2 ^1A_1$ 
excited-state curve.

The only state which shows triplet instabilities is the $1^3B_2$ state.  The TDLDA  $1^3B_2$
energies are {\em imaginary} (even if designated as negative in Figs.~\ref{fig:all_TDDFT_curves}
and \ref{fig:B2_curves}) 
between about about 100$^\circ$ and about 140$^\circ$ while the TDB3LYP  $1^3B_2$ energies are 
{\em imaginary}
over a larger range, between about 90$^\circ$ and 160$^\circ$.  Certainly, one way to understand
how this can happen is through the formulae Eq.~(\ref{eq:results.3}), but a better way to
understand triplet instabilities is in terms of the fact that the quality of response theory
energies depends upon having a high-quality ground state.  Problems occur in TDDFT excitation
energies because the functional $E_{xc}$ is only approximate.  

Stability analysis and triplet instabilities
have already been discussed in a general way in Sec.~\ref{sec:TDDFT} where it was seen 
that no symmetry breaking is expected for a closed-shell singlet ground state when $E_{xc}$ is exact.
However, symmetry breaking can occur when $E_{xc}$ is approximate.  To deepen our 
understanding of triplet instabilities, let us look at a two-orbital model for the 
dissociation of the $\sigma$ bond in H$_2$, which is 
similar to the dissociation of the CC $\sigma$ bond in oxirane.  Here we follow the argument
in Ref.~\cite{CGG+00} and consider the wave function,
\begin{equation}
  \Psi_\lambda = \vert \sqrt{1-\lambda^2} \sigma + \lambda \sigma^* , \sqrt{1-\lambda^2} \bar{\sigma}
  - \lambda \bar{\sigma}^* \vert \,,
  \label{eq:results.6}
\end{equation}
which becomes $\vert s_A, \bar{s}_B \vert$ as $\lambda \rightarrow 1/\sqrt{2}$.
The corresponding energy expression is,
\begin{equation}
  E_\lambda = E_0 + 2\lambda^2 \frac{\omega_T^2}{\epsilon_{\sigma^*} - \epsilon_\sigma} + {\cal O}(\lambda^3) \, .
  \label{eq:results.7}
\end{equation}
This result suggests the more general result \cite{C01} that symmetry breaking will occur if and only if there 
is an imaginary triplet excitation energy ($\omega_T^2<0$).  Certainly, one way to overcome the problem
of triplet excitation energies is to improve the quality of the exchange-correlation functional.
However, another way is to use the TDA.  The reason is
clear in HF theory where TDHF/TDA is the same as CIS, whose
excited states are a rigorous upper bound to the true excitation energies.  
Therefore, not only will the excitations remain real but variational
collapse will not occur.  To our knowledge, there is no way to extend these ideas to justify the use of the
TDA in the context of TDDFT calculations except by carrying out explicit calculations to show that the TDA
yields improved PESs for TDDFT calculations.  This was previously shown for H$_2$ \cite{CGG+00} and 
is evidently also true for oxirane judging from Fig.~\ref{fig:B2_curves} where the TDLDA/TDA
curve is remarkably similar to the DMC curve.

As shown in Fig.~\ref{fig:B2_curves}, triplet instabilities are often also associated with 
singlet near instabilities.  In this case, the TDDFT $1 ^1B_2$ singlet excitation energies
are much too low.  The TDA brings the TDLDA curve into the same energetic region of
space as the corresponding DMC curve, but still compared with the DMC curve 
the TDLDA TDA curve appears to be qualitatively incorrect after 120$^\circ$ 
where it is seen to be decreasing, rather than increasing, in energy as the
angle opens.  This, in fact, is the behavior observed in the CAS(4,6) singlet
calculation but is opposite to what happens in the more accurate DMC calculation.

\begin{figure}
   \includegraphics[width=\columnwidth]{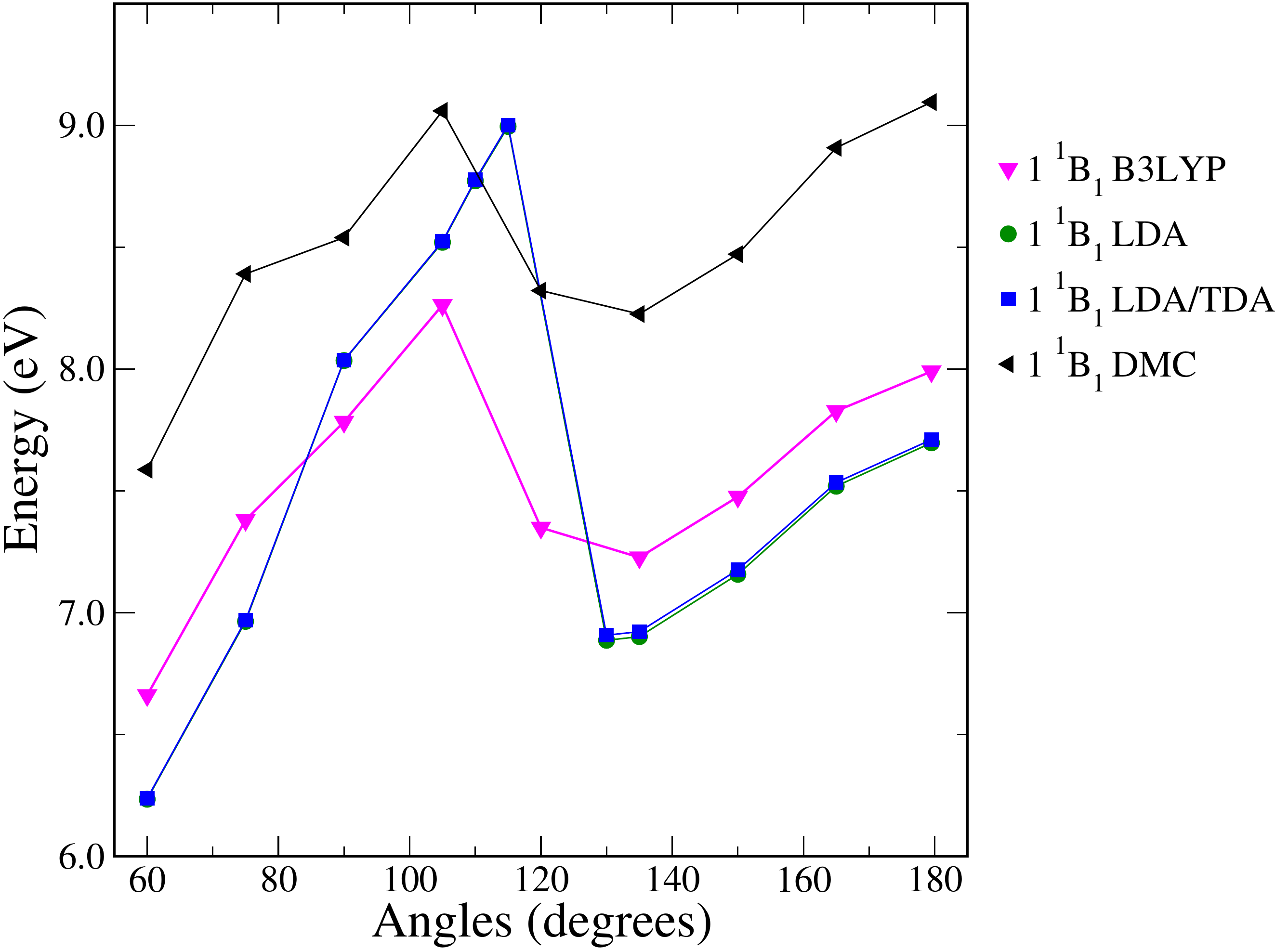}
   \caption{$C_{2v}$ ring opening curves: $1^1B_1$ state. The TDLDA and TDLDA/TDA
   curves are practically superimposed.}
  \label{fig:1B1_curves}
\end{figure}

\begin{figure*}[t]
   \includegraphics[scale=0.32]{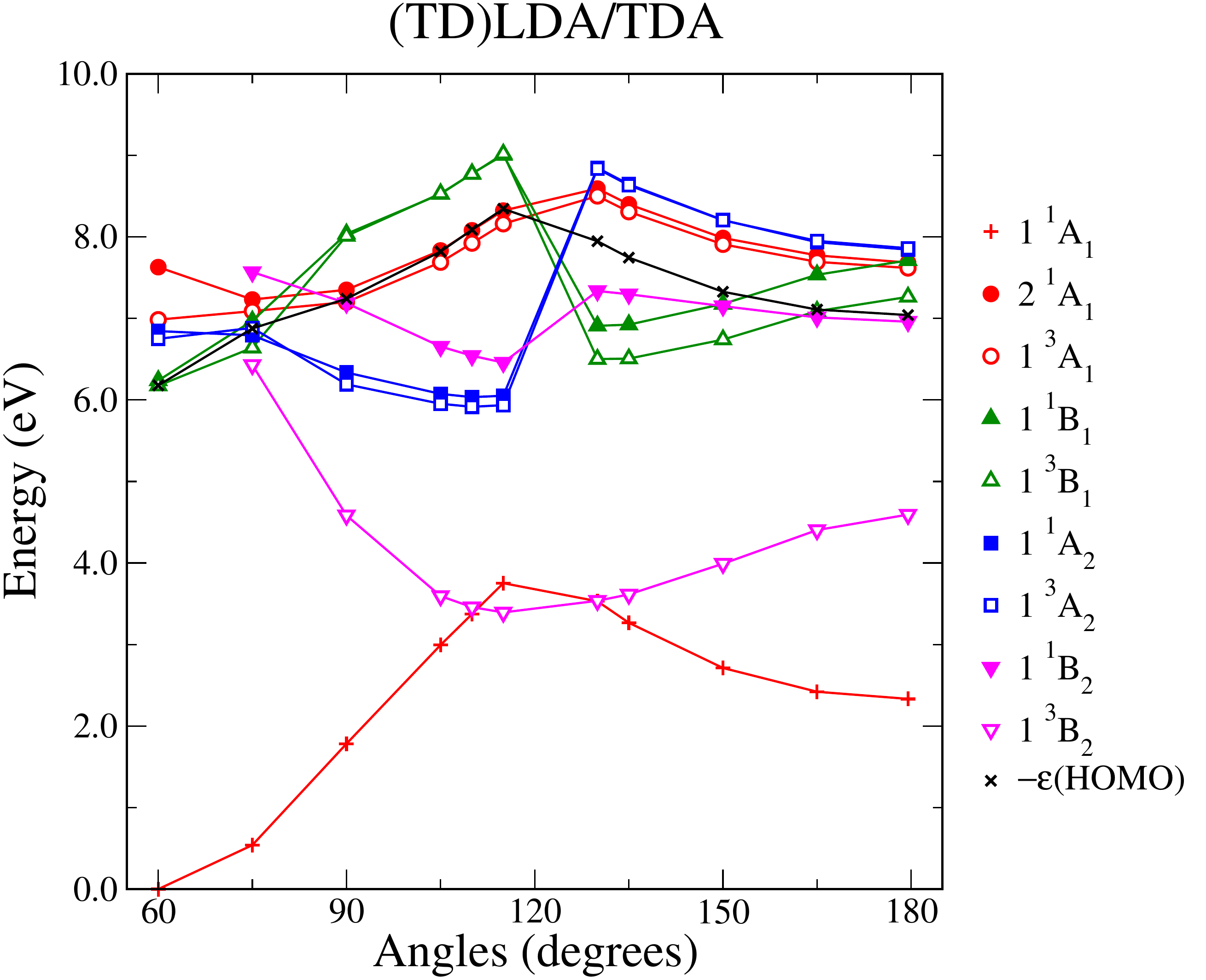}
   \includegraphics[scale=0.32]{fig7b.pdf}
   \caption{Comparison of TDLDA/TDA and DMC $C_{2v}$ ring opening curves.}
  \label{fig:DMC_and_TDA_curves}
\end{figure*}
The remaining $1^{1,3}B_1$ and $1^{1,3}A_2$ states are primarily Rydberg in nature with
the corresponding singlets and triplets being energetically nearly degenerate.  The graph
for a single one of these states suffices to illustrate the general trend observed in the
case of all four states.  Fig.~\ref{fig:1B1_curves} shows what happens for the $1^1B_1$
states.  All the curves have qualitatively the same form.  This is particularly true for
the TDB3LYP curve which closely resembles the DMC curve but shifted down by a little over
one eV in energy.  Notice also that the TDLDA and TDLDA/TDA curves are essentially superimposed.
This is because,
\begin{eqnarray}
  \left(\omega_S\right)_{TDA}^2 - \omega_S^2 & = & (ia \vert 2f_H + f_{xc}^{\uparrow,\uparrow}
    + f_{xc}^{\uparrow,\downarrow} \vert ai )^2 \nonumber \\
  \left(\omega_T\right)_{TDA}^2 - \omega_T^2 & = & (ia \vert f_{xc}^{\uparrow,\uparrow}
    - f_{xc}^{\uparrow,\downarrow} \vert ai )^2 \, .
  \label{eq:results.8}
\end{eqnarray}
For Rydberg states the overlap between orbitals $i$ and $a$ is very small and so the difference
between full and TDA TDLDA calculations can be neglected.  From Eq.~(\ref{eq:results.2}) this
also means that the Rydberg excitation energies also reduce to simple orbital energy differences.

Fig.~\ref{fig:DMC_and_TDA_curves} shows that the TDLDA/TDA yields, at least in
this case, a mostly qualitatively reasonable description of photochemically important PESs
as compared with high-quality QMC PESs.
By this we only mean that corresponding TDLDA/TDA and QMC potential energy curves
have typically the same overall form and are found at roughly similar energies.
As discussed above, significant differences do remain in the shapes of some potential 
energy curves ($2 ^1A_1$ and $1 ^1B_2$) and the detailed energy ordering of the curves
differs between the TDLDA/TDA and QMC results.

\subsection{{\em Con}- and {\em Dis}rotatory Ring Opening}

\begin{figure}[b]
   \includegraphics[angle=0,scale=0.50]{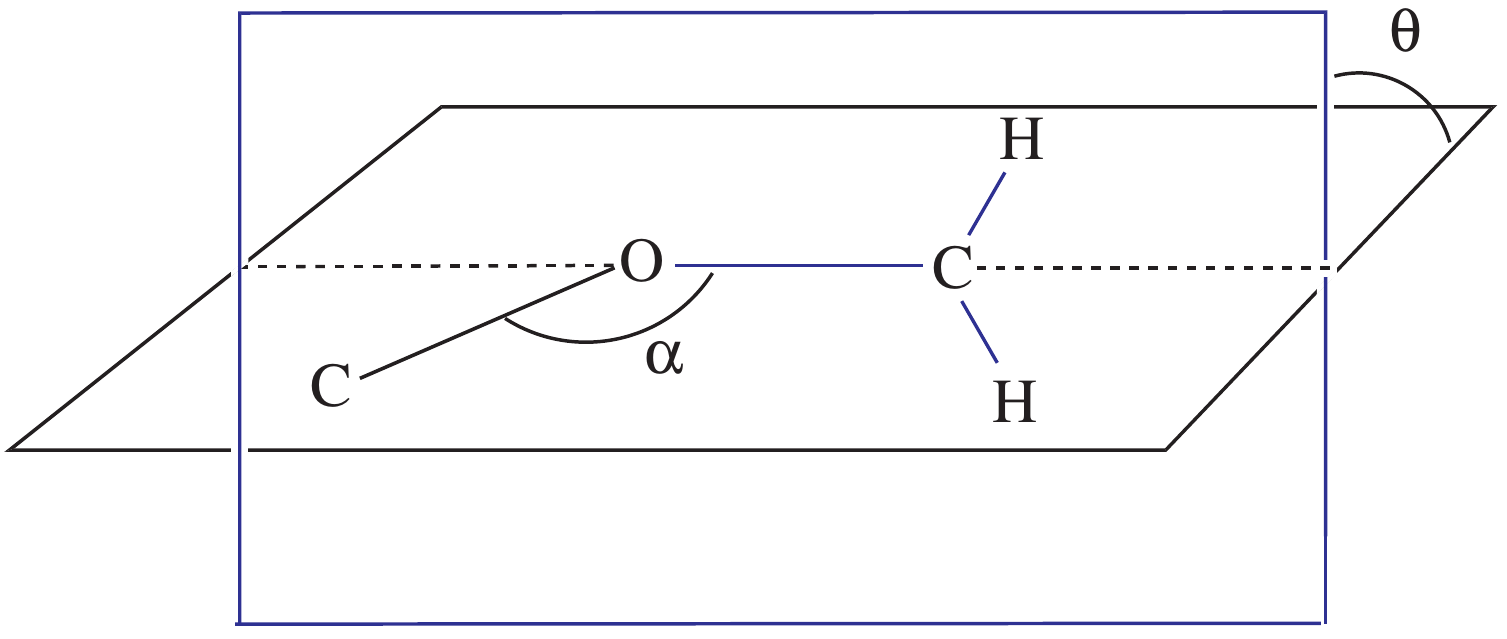}
  \caption{$(\alpha,\theta)$ coordinate system used in this paper.}
  \label{fig:coord}
\end{figure}

Conrotatory and disrotatory potential energy surfaces were calculated using the
simplifying assumption that each set of 4 atoms OCH$_2$ is constrained to lie
in a plane.  Our coordinate system is defined in Fig.~\ref{fig:coord}.  All other
geometrical parameters were relaxed for the thermal reaction preserving $C_s$
and $C_2$ symmetry for the {\em con} and the {\em dis} path, respectively.

\subsubsection{Thermal Reaction}

\begin{figure}[t]
 \includegraphics[angle=0,scale=0.55]{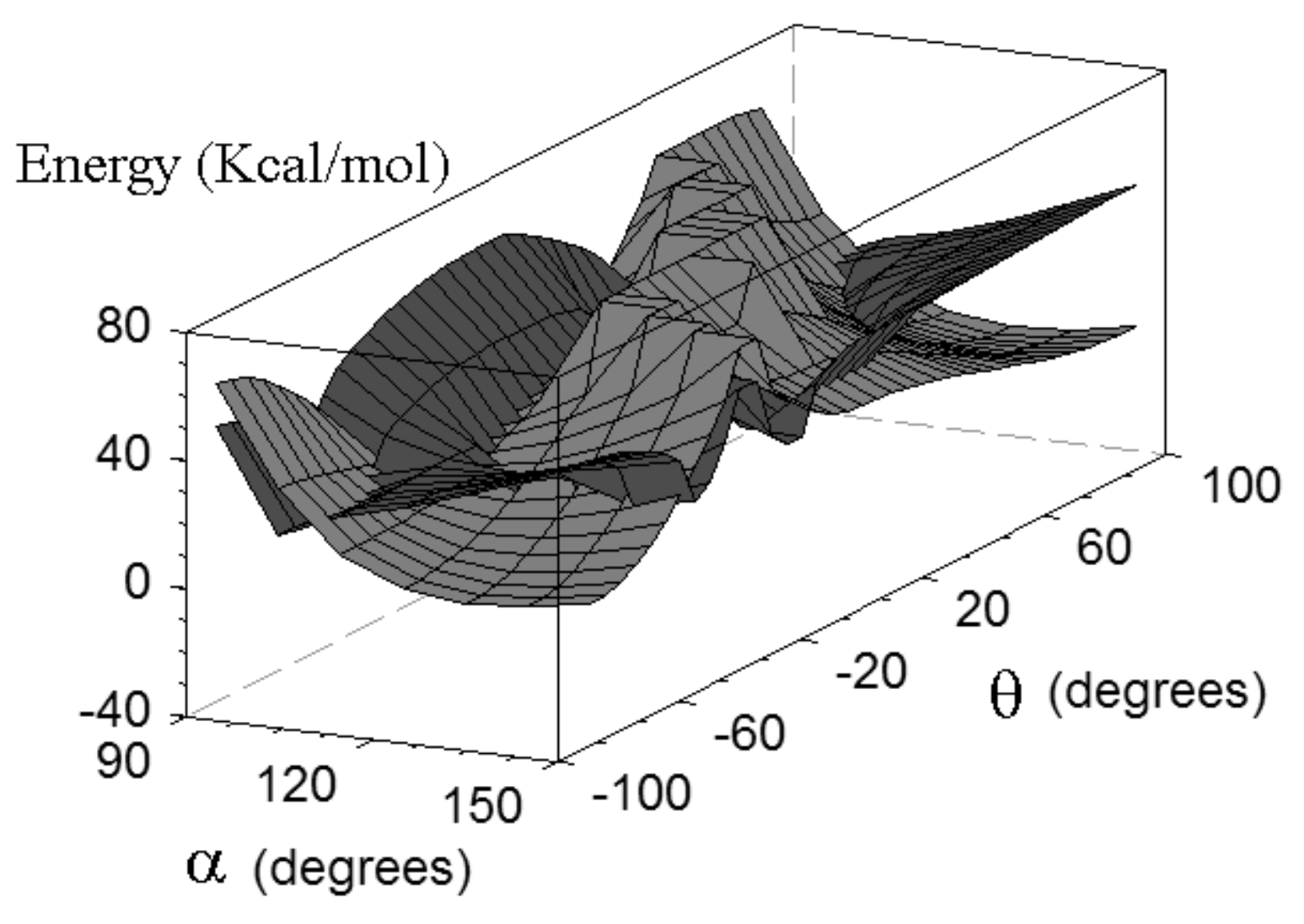}
 \caption{TDLDA $S_0$ (light grey) and $T_1$ (dark grey) PESs.  Note that a 
 ``negative excitation energy'' is just a convenient graphical trick for representing 
 an imaginary excitation energy.  Negative excitation
 energies correspond to triplet instabilities.}
\label{fig:TDLDA_surfaces}
\end{figure}

\begin{figure}
 \includegraphics[angle=0,scale=0.55]{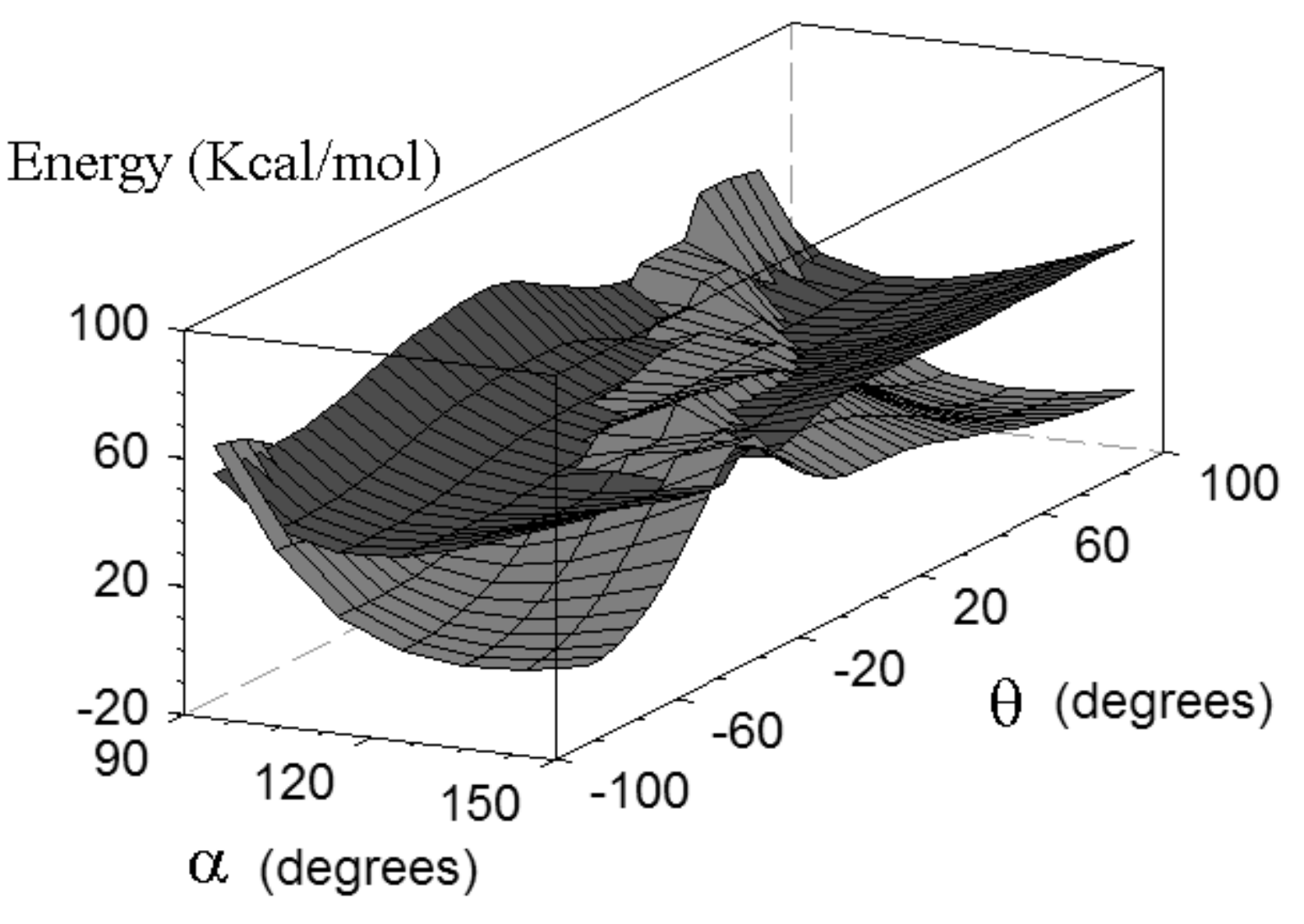}
 \caption{TDLDA/TDA $S_0$ (light grey) and $T_1$ (dark grey) PESs.  In this
 case negative excitation energies are real, not imaginary, quantities.}
\label{fig:TDA_surfaces}
\end{figure}

Fig.~\ref{fig:TDLDA_surfaces} shows the ground state ($S_0$) PES.  In accordance with 
the WH model there appears to be a much lower energy barrier for {\em con} ring opening
than for {\em dis} ring opening.  Also shown is the lowest triplet state ($T_1$).  It is
clear that triplet instabilities occur along the disrotatory pathway as the CC $\sigma$
bond breaks.  Remarkably, they do not occur along the conrotatory pathway in the LDA.
(They should, of course, be absent when the exchange-correlation functional
is exact.) Although triplet instabilities account for about 50\% of the LDA surface,
this fraction actually increases to 93\% for the B3LYP functional where triplet instabilities
occur along both the conrotatory and disrotatory reaction pathways.  Finally, when
the HF method is used triplet instabilities account for  100\% of the PES.
Perhaps because we forced the OCH$_2$ to lie in a single plane, triplet instabilities
appear to be nearly everywhere in these last methods, potentially spelling trouble for
mixed TDDFT/classical photodynamics calculations.
We would thus like to caution against the use of hybrid functionals for this type
of application.  At the same time, we reiterate our recommendation to use the TDA
because, of course, the TDLDA/TDA excited state PESs
(Fig.~\ref{fig:TDA_surfaces}) show no triplet instabilities.

\subsubsection{Photochemical Reaction}

\begin{figure}
 \includegraphics[angle=0,scale=0.55]{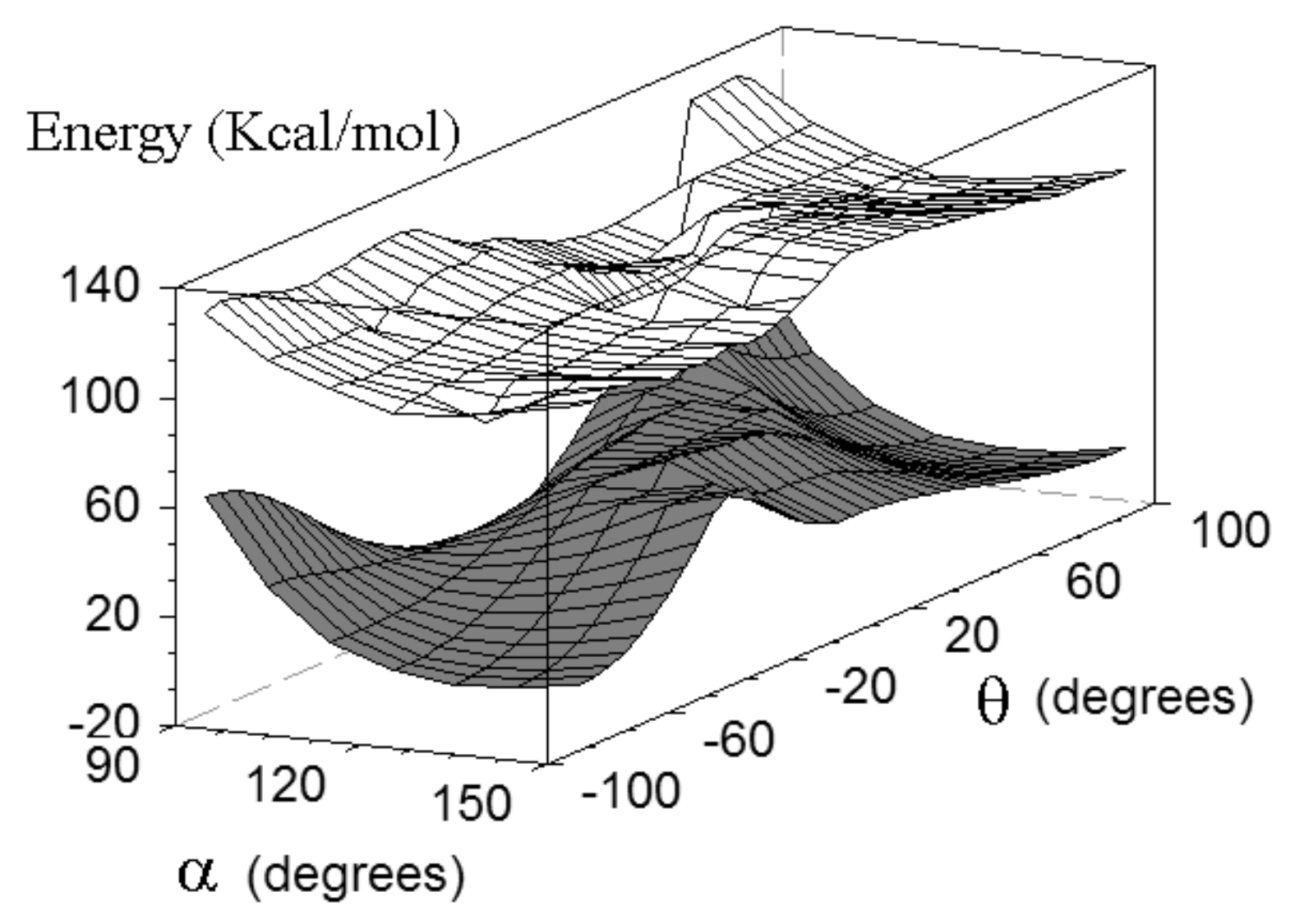}
 \caption{TDLDA/TDA $S_0$ (light grey) and $S_1$ (white) potential energy surfaces.}
 \label{fig:TDA_singlet_surfaces}
\end{figure}

In considering whether the WH model has any validity for describing con- versus
disrotatory rotation for photochemical ring opening in oxirane, we are immediately
faced with the problem that there are a large number of excited states with similar
energies which cross each other (Fig.~\ref{fig:CAS_and_DMC_curves}).  
Provisionally, we have decided to assume that state symmetry is conserved
and so to look only at one PES, namely the $1 ^1B_1 S_1$ surface (which we shall 
simply refer to as $S_1$) which begins as the $^1[2b_1(n)\rightarrow 7a_1(3s)]$ 
excitation for the closed cycle and then evolves as the molecular geometry changes.  
In so doing, we are following the spirit of simple WH theory which makes maximum
use of symmetry.  Turning back to this theory
(Fig.~\ref{fig:WHcorrdiag}) and thinking of excitation to a Rydberg state as analogous 
to electron removal, it suffices to remove an electron from the $n$ orbital of the
{\em thermal} WH diagram, thereby giving the $\sigma^2 n^1$ configuration, 
to have an idea of what might be the relative importance of the
{\em con} and {\em dis} mechanisms for the $S_1$ surface.  
In this case the {\em con} mechanism leads to the energetically nearest-neighbor 
single excitation, $\sigma \rightarrow n$, and the {\em dis} mechanism leads to the 
nearest-neighbor single excitation, $\sigma\rightarrow \sigma^*$.
Since both mechanisms correspond to nearest-neighbor excitation, 
no particular preference is {\em a priori} expected for one mechanism over the other.  
The ground ($S_0$) and excited state singlet ($S_1$) surfaces calculated with the TDLDA/TDA 
are shown in Fig.~\ref{fig:TDA_singlet_surfaces}.  Indeed the simple WH theory appears 
to be confirmed in that the $S_1$ surface is remarkably flat.

Unfortunately this analysis is far too simplistic.   In particular, Kasha's rule \cite{K50} 
tells us that the first excited triplet ($T_1$) or singlet ($S_1$) states are the most 
likely candidates for the initiation of a photochemical reaction,
where now $T_1$ and $S_1$ are the {\em globally} lowest states and not necessarily 
the lowest state of a given symmetry.
This is based upon the idea that relaxation of higher excited states is rapid.
Such relaxation is due to environmental effects or vibronic coupling which need not
preserve the symmetry of the electronic state.  Hence $S_1$ at one geometry 
is not necessarily $S_1$ at another geometry.
Indeed looking again at the DMC curves in 
Fig.~\ref{fig:CAS_and_DMC_curves}, it is easy to believe that the molecule will
eventually arrive in the $1 ^1B_2(\sigma,\sigma^*)$ state during ring opening and
that, because the $\sigma$ orbital is higher in energy than the $\sigma^*$ orbital
at this geometry (Fig.~\ref{fig:Walsh_C2v}), that the usual WH argument will 
still predict a preference for the {\em dis} mechanism. 
 
Again we are falling into a trap imposed by the use of symmetry.  Excitation, 
for example, from the $2b_1(n)$ orbital (Fig.~\ref{fig:orbitals}) into 
the $8a_1(3p_z)$ Rydberg orbital (Fig.~\ref{fig:excite_orbs}) might lead to 
preferential CO, rather than CC, bond breaking to the extent that ring opening
augments the valence CO($\sigma*$) character of the target orbital \cite{T07}.
The difficulty of predicting {\em a priori} such behavior is part of what 
motivates us to move towards dynamics as a better tool for photochemical modeling.

\section{Conclusion}
\label{sec:conclude}

In this paper, we have examined the potential energy curves and surfaces
for the symmetric ring opening of oxirane to assess possible
difficulties which might be encountered when using TDDFT in photodynamics
simulations of its photochemistry.  Our TDDFT calculations provided
useful insight helpful in constructing the active space needed for
more accurate CASSCF and still more accurate DMC calculations.  
It is indeed worth noting that identifying active
spaces is probably one of the more common uses of TDDFT in photochemical
studies.  Here, we summarize our main conclusions obtained by comparing 
our TDDFT and DMC results.

Oxirane does not seem to be a molecule where charge transfer excitations 
are important.  The artificially low ionization threshold typical in TDDFT 
for most functionals is still high enough for the B3LYP functional so as not 
to pose a serious problem.  Even for the LDA functional, where the TDDFT 
ionization threshold is lower, the shapes of the excited-state Rydberg curves 
seem to be qualitatively correct even if they should not be considered 
quantitative.  

Problems do show up as the CC $\sigma$ bond breaks.  By far, the most 
severe problem encountered is the presence of triplet instabilities and 
singlet near instabilities, where the excited-state PES takes an unphysical 
dive in energy towards the ground state.  In the case of the triplet, the excitation
energy may even become imaginary, indicating that the ground state energy
could be further lowered by allowing the Kohn-Sham orbitals to break symmetry.
Although this problem is very much diminished compared to that seen in the 
HF ground state, it is still very much a problem as seen for example in
the 2D surfaces of the con and disrotatory ring opening of oxirane.  

It is difficult to overstate the gravity of the triplet instability problem
for the use of TDDFT in photodynamics simulations.
One should not allow symmetry breaking as it is an artifact which should not 
occur when the exchange-correlation functional is exact. Moreover, there may be more 
than one way to lower the molecular energy by breaking symmetry, and searching 
for the lowest energy symmetry broken solution can be time consuming.
Finally, the assignment of excited states using 
broken symmetry orbitals is far from evident.  On the whole, it thus seems
better to avoid the problem by using the TDA to decouple (at least partially)
the quality of the excited-state PES from that of the ground state PES.
Our calculations show that this works remarkably well for the $1^3B_2$
curve along the $C_{2v}$ pathway, in comparison with good DMC results.
The $1^1B_2$ state, which appeared to collapse in energy without the TDA
in the region of bond breaking, is also restored to a more reasonable range
of energies.  For this reason, we cannot even imagine carrying out photochemical
simulations with TDDFT without the use of the TDA.  Or, to put it more bluntly,
the TDA, which is often regarded as an approximation on conventional TDDFT 
calculations, gives {\em better} results than does conventional TDDFT when
it comes to excited-states PESs in situations where bond breaking occurs.

The TDA is unable to solve another problem which occurs as the CC $\sigma$ bond
breaks, namely the presence of an unphysical cusp on the ground state curve
(or surface).  This cusp is there because of the difficulty of approximate
density functionals to describe a biradical structure with a single determinantal
wave function.  The cusp is also translated up onto the excited-state curves 
because of the way that PESs are constructed in TDDFT.  However the problem
is very much diminished compared to that seen in HF because of the presence
of some correlation in DFT even when the exchange-correlation functional is 
approximate.  The ultimate solution to the cusp problem is most likely some 
sort of explicit incorporation into TDDFT of two- and higher-electron 
excitations.  Among the methods proposed for doing just this are dressed TDDFT
or polarization propagator corrections \cite{CZMB04,MZCB04,C05} and 
spin-flip TDDFT \cite{SHK03,SK03,WZ04}.

Our examination of the con- and disrotatory ring-opening pathways revealed
another important point which has nothing to do with TDDFT.  This is that
the manifold of excited state PESs is too complicated to interpret easily.
The best way around this is to move from a PES interpretation 
to a pathway interpretation of photochemistry, and the best 
way to find the pathways moving along and between adiabatic PESs seems 
to be to carry out dynamics.
For now, it looks as though the use of TDDFT/TDA in photodynamics simulations
may suffice to describe the principal photochemical processes in oxirane.
On-going calculations \cite{T07} do indeed seem to confirm this assertion in 
so far as trajectories have been found in good agreement with the accepted 
Gomer-Noyes mechanism \cite{GN50,IIT73}. 

\appendix
\section{Photochemistry of Oxiranes}
\label{sec:photochem}

\begin{figure}
 \includegraphics[angle=0,scale=0.4]{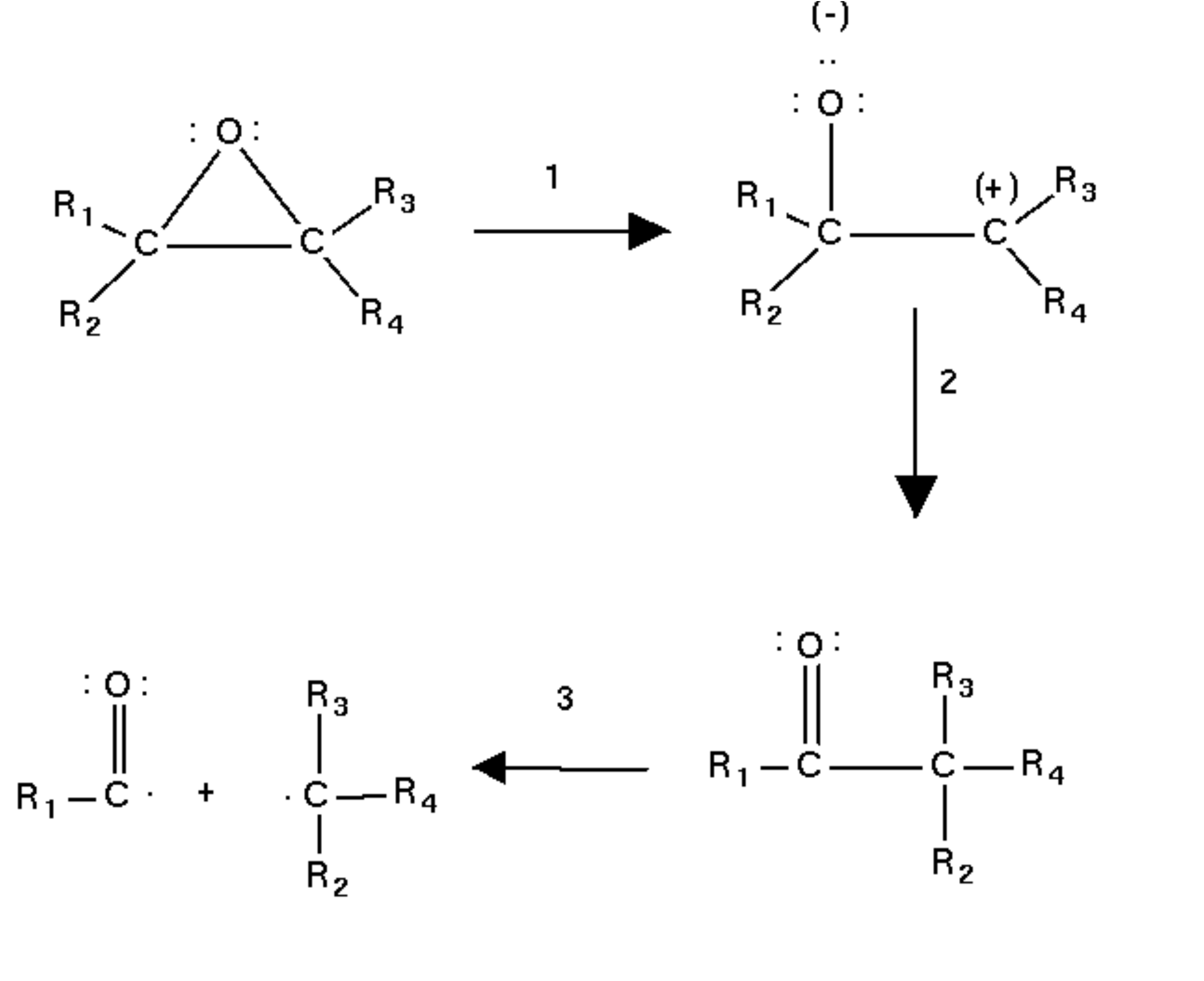}
 \caption{Typical reactions of alkyl oxiranes.}
 \label{fig:alkyl_oxirane}
\end{figure}

\begin{figure}
 \includegraphics[angle=0,scale=0.3]{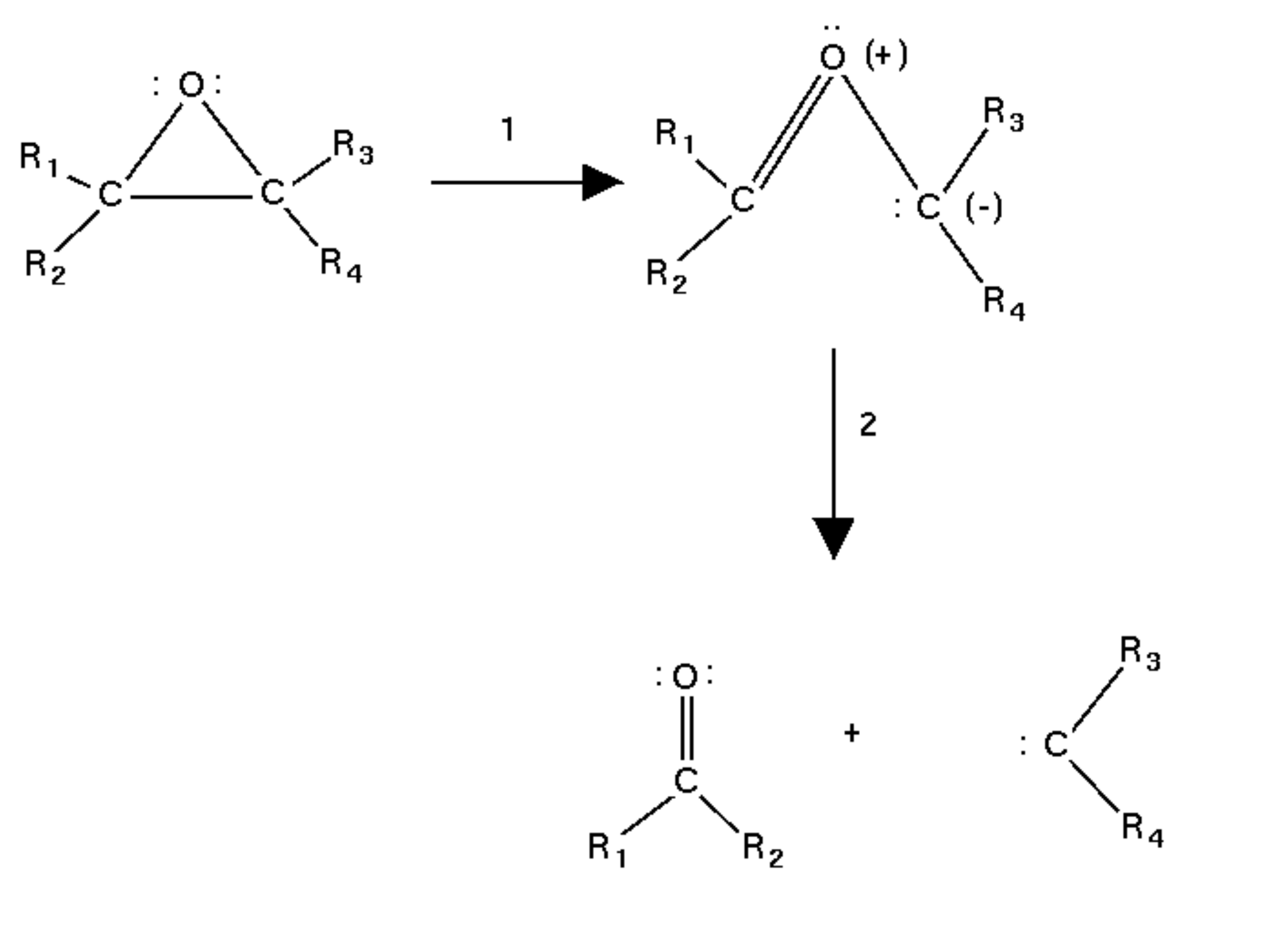}
 \caption{Typical reactions of aryl oxiranes.}
\label{fig:aryl_oxirane}
\end{figure}

The generic structure of oxiranes is shown as the starting point in the
chemical reactions in Figs.~\ref{fig:alkyl_oxirane} and \ref{fig:aryl_oxirane}.
When R$_1$, R$_2$, R$_3$, and R$_4$ are hydrogens or alkyl groups, then the prefered reaction 
is CO cleavage both photochemically and thermally (Step 1, Fig.~\ref{fig:alkyl_oxirane}).  
In particular it is estimated that the CO rupture energy is 
about 52 kcal/mol while the CC rupture energy is 5-7 kcal/mol higher \cite{BSD79a}.  
Since the molecule is not symmetric along the CO ring-opening pathway, the WH model does
not apply.  Photochemical
CO ring opening may be followed by alkyl migration \cite{AWR+04} (Step 2, Fig.~\ref{fig:alkyl_oxirane}).
In the particular case of oxirane itself (R$_1$=R$_2$=R$_3$=R$_4$=H), hydrogen migration is followed
by breaking of the CC single bond (Step 3, Fig.~\ref{fig:alkyl_oxirane}).  This is the Gomer-Noyes
mechanism \cite{GN50} which was confirmed experimentally by Ibuki, Inasaki, and Takesaki 
\cite{IIT73}.

In contrast, cyano and aryl substitutions favor CC bond breaking 
(Step 1, Fig.~\ref{fig:aryl_oxirane}) to form
what is often refered to as a 1,3-dipolar species or a carbonyl ylide.  This is the case where there
may be sufficient symmetry that the WH model applies.  
The photochemistry of phenyl and phenyl substituted oxiranes has been reviewed in 
Refs.~\cite{H77,HRS+80,P87} and on pages 565-566 of Ref.~\cite{T91}.
Evidence for carbonyl ylides
goes back to at least the 1960s when Ullman and Milks investigated the tautomerization of 2,3-diphenylindenone oxide
\cite{UM62,UM64} and Linn and Benson investigated the ring-opening reaction of tetracyanoethylene oxide
\cite{LB65,L65}.  Finding cases where the WH model applies and where its predictions can be verified
turns out to be not straightforward, particularly in the 
photochemical case.  There are several reasons for this.  First of all, too much asymmetry should be
avoided in order to assure the applicability of the WH orbital symmetry conservation rule and so
avoid passing directly onto carbene formation (Step 2, Fig.~\ref{fig:aryl_oxirane}).  Secondly,
the substituted oxirane should include groups which are bulky enough to confer the structural 
rigidity needed to avoid premature radiationless relaxation, but not bulky enough to favor
carbene formation.  The predictions of the WH model have been found to hold
for {\em cis}- and {\em trans}-1,2-diphenyloxirane \cite{MP84} while carbene formation dominates
for tetraphenyl oxirane \cite{TABG80}.

\section{Supplementary Material}
\label{sec:suppl}

Benchmark quality diffusion Monte Carlo (DMC) energies 
calculated at the geometries given in  Table~\ref{tab:B3LYPopt}
are reported 
here for the ground state ($1 ^1A_1$, Tables~\ref{tab:A1} and \ref{tab:SA-A1}) 
and the lowest excited state of each symmetry ($2 ^1A_1$, Table~\ref{tab:SA-A1};
$1 ^3A_1$,  Table~\ref{tab:A1}; $1 ^1B_1$, $1 ^3B_1$, Table~\ref{tab:B1};
$1 ^1A_2$, $2 ^3A_2$, Table~\ref{tab:A2}; and $1 ^1B_2$, $1 ^3B_2$, 
Table~\ref{tab:B2}).  We set the zero of the energy to coincide with
the DMC $1 ^1A_1$ energy at 60$^\circ$ and report
the statistical error on the energy in parenthesis.
Note that the negative value of the $1 ^1A_1$ SA-DMC energy at 
60$^\circ$ is statistically compatible with zero.
We hope that these DMC data will encourage further
developing and testing of improved TDDFT algorithms suitable for
addressing the problems mentioned in this article.

\begin{table}
\caption{$A_1$ DMC energies as a function of COC ring-opening angle
along the $C_{2v}$ ring-opening pathway.}
\label{tab:A1}
\begin{tabular}{ccc}
\hline \hline 
\\
\multicolumn{3}{c}{$A_1$ DMC Energies and statistical error in Hartree}  \\
\\
$\angle$ COC ($^\circ$)  & $1 ^1A_1$ & $1 ^3A_1$ \\
\hline
60.   &   0.0000 (0.0009) &   0.3099 (0.0010) \\
75.   &   0.0212 (0.0009) &   0.2811 (0.0010) \\
90.   &   0.0651 (0.0010) &   0.2836 (0.0009) \\
105.  &   0.0998 (0.0010) &   0.2998 (0.0010) \\ 
120.  &   0.1222 (0.0009) &   0.3406 (0.0011) \\
135.  &   0.1153 (0.0009) &   0.3223 (0.0009) \\
150.  &   0.1083 (0.0009) &   0.3106 (0.0010) \\
165.  &   0.1048 (0.0010) &   0.3025 (0.0010) \\
179.5 &   0.1017 (0.0010) &   0.3007 (0.0010) \\
\\
\hline \hline
\end{tabular}
\end{table}
\begin{table}
\caption{$^1A_1$ SA-DMC energies as a function of COC ring-opening angle
along the $C_{2v}$ ring-opening pathway.}
\label{tab:SA-A1}
\begin{tabular}{ccc}
\hline \hline 
\\
\multicolumn{3}{c}{$^1A_1$ SA-DMC Energies and statistical error in Hartree}  \\
\\
$\angle$ COC ($^\circ$)  & $1 ^1A_1$ & $2 ^1A_1$ \\
\hline
60.   &  -0.0018 (0.0009) &   0.3246 (0.0009) \\
75.   &   0.0254 (0.0010) &   0.2904 (0.0009) \\
90.   &   0.0694 (0.0010) &   0.2865 (0.0010) \\
105.  &   0.1011 (0.0010) &   0.2962 (0.0009) \\ 
120.  &   0.1263 (0.0009) &   0.2699 (0.0010) \\
135.  &   0.1196 (0.0010) &   0.2902 (0.0010) \\
150.  &   0.1103 (0.0010) &   0.3109 (0.0011) \\
165.  &   0.1054 (0.0010) &   0.3063 (0.0010) \\
179.5 &   0.1037 (0.0010) &   0.3054 (0.0010) \\
\\
\hline \hline
\end{tabular}
\end{table}
\begin{table}
\caption{$B_1$ DMC energies as a function of COC ring-opening angle
along the $C_{2v}$ ring-opening pathway.}
\label{tab:B1}
\begin{tabular}{ccc}
\hline \hline 
\\
\multicolumn{3}{c}{$B_1$ DMC Energies and statistical error in Hartree}  \\
\\
$\angle$ COC ($^\circ$)  & $1 ^1B_1$ & $1 ^3B_1$ \\
\hline
60.   &   0.2788 (0.0010) &   0.2636 (0.0010) \\
75.   &   0.3083 (0.0009) &   0.3026 (0.0010) \\
90.   &   0.3138 (0.0010) &   0.3140 (0.0009) \\
105.  &   0.3329 (0.0010) &   0.3316 (0.0010) \\ 
120.  &   0.3058 (0.0010) &   0.2970 (0.0010) \\
135.  &   0.3023 (0.0010) &   0.2887 (0.0010) \\
150.  &   0.3113 (0.0010) &   0.2972 (0.0010) \\
165.  &   0.3274 (0.0010) &   0.3111 (0.0011) \\
179.5 &   0.3343 (0.0010) &   0.3155 (0.0010) \\
\\
\hline \hline
\end{tabular}
\end{table}
\begin{table}
\caption{$A_2$ DMC energies as a function of COC ring-opening angle
along the $C_{2v}$ ring-opening pathway.}
\label{tab:A2}
\begin{tabular}{ccc}
\hline \hline 
\\
\multicolumn{3}{c}{$A_2$ DMC Energies and statistical error in Hartree}  \\
\\
$\angle$ COC ($^\circ$)  & $1 ^1A_2$ & $1 ^3A_2$ \\
\hline
60.   &   0.2972 (0.0010) &   0.3001 (0.0009) \\
75.   &   0.3021 (0.0010) &   0.2971 (0.0009) \\
90.   &   0.2901 (0.0010) &   0.2857 (0.0010) \\
105.  &   0.2835 (0.0010) &   0.2769 (0.0010) \\ 
120.  &   0.2870 (0.0010) &   0.2843 (0.0010) \\
135.  &   0.3344 (0.0010) &   0.3319 (0.0011) \\
150.  &   0.3179 (0.0010) &   0.3165 (0.0010) \\
165.  &   0.3111 (0.0010) &   0.3088 (0.0010) \\
179.5 &   0.3086 (0.0010) &   0.3066 (0.0010) \\
\\
\hline \hline
\end{tabular}
\end{table}
\begin{table}
\caption{$B_2$ DMC energies as a function of COC ring-opening angle
along the $C_{2v}$ ring-opening pathway.}
\label{tab:B2}
\begin{tabular}{ccc}
\hline \hline 
\\
\multicolumn{3}{c}{$B_2$ DMC Energies and statistical error in Hartree}  \\
\\
$\angle$ COC ($^\circ$)  & $1 ^1B_2$ & $1 ^3B_2$ \\
\hline
60.   &   0.3484 (0.0009) &   0.3301 (0.0010) \\
75.   &   0.3040 (0.0010) &   0.2395 (0.0010) \\
90.   &   0.2810 (0.0010) &   0.1657 (0.0010) \\
105.  &   0.2549 (0.0011) &   0.1229 (0.0009) \\ 
120.  &   0.2470 (0.0011) &   0.1165 (0.0010) \\
135.  &   0.2481 (0.0010) &   0.1232 (0.0009) \\
150.  &   0.2604 (0.0010) &   0.1410 (0.0010) \\
165.  &   0.2718 (0.0010) &   0.1611 (0.0009) \\
179.5 &   0.2733 (0.0010) &   0.1667 (0.0010) \\
\\
\hline \hline
\end{tabular}
\end{table}
\begin{table}
\caption{Oxirane $C_{2v}$ geometries obtained at different
COC ring opening angles with all other parameters optimized at the
B3LYP level.}
\label{tab:B3LYPopt}
\begin{tabular}{ccccc}
\hline \hline 
\\
\multicolumn{5}{c}{$C_{2v}$ geometries}  \\
\\
$\angle$ COC ($^\circ$)  & 
$\angle$ HCH ($^\circ$)  & 
$\angle$ HCOC ($^\circ$)  & 
$R(\mbox{CH})$ (\AA) &  
$R(\mbox{CO})$ (\AA) \\
\hline
60.           &   115.52      &     111.317    & 1.08375  &   1.44558  \\
75.           &   118.69      &     105.214    & 1.08311  &   1.36333  \\
90.           &   122.04      &      97.339    & 1.08131  &   1.34639  \\
105.          &   122.28      &      84.096    & 1.08311  &   1.35811  \\
120.          &   118.06      &     109.402    & 1.08576  &   1.38032  \\
135.          &   119.88      &     105.829    & 1.08353  &   1.33439  \\
150.          &   121.78      &     101.409    & 1.08154  &   1.30455  \\
165.          &   123.27      &      96.054    & 1.08009  &   1.28694  \\
179.5         &   123.86      &      90.207    & 1.07954  &   1.28114  \\
\\
\hline \hline
\end{tabular}
\end{table}

\begin{acknowledgments}

This study was carried out in the context of a Franco-Mexican collaboration financed through
{\em ECOS-Nord} Action M02P03.  
Financing by {\em ECOS-Nord} made possible working visits of AV to the {\em Universit\'e 
Joseph Fourier} and of MEC to {\em Cinvestav}.  FC acknowledges support 
from the Mexican Ministry of Education via a CONACYT (SFERE 2004) scholarship and from 
the {\em Universidad de las Americas Puebla} (UDLAP).  

We like to acknowledge useful discussions with  Mathieu Maurin,  
Enrico Tapavicza, Neepa Maitra, Ivano Tavernelli, Todd Mart\'inez, and Massimo
Olivucci.

MEC and FC thank Pierre Vatton, Denis Charapoff, R\'egis Gras, 
S\'ebastien Morin, and Marie-Louise Dheu-Andries for technical support of 
the {\em D\'epartement de Chimie Mol\'ecularie} and 
{\em Centre d'Exp\'erimentation le Calcul Intensif en Chimie} (CECIC) computers.
CF aknowledges the support by the {\em Stichting Nationale Computerfaciliteiten} 
(NCF-NWO) for the use of the SARA supercomputer facilities.

\end{acknowledgments}


\begin{thebibliography}{999}
\bibitem{symmetry} ``... it is probably fair to say that in 
photochemical reactions, the unsymmetrical reaction paths are the ones most 
normally followed.  In spite of this, most illustrations of potential energy 
curves in the literature and in this book are for symmetrical paths, 
since these are much easier to calculate or guess.  It is up to
the reader of any of the theoretical photochemistry literature to keep this in
mind and to correct for it the best he or she can ...'' (Ref.~\cite{KM95}, p.\ 218.)
\end{thebibliography}
\end{document}